\documentclass[10pt, lettersize,journal]{IEEEtran}
\usepackage{setspace} 
\usepackage{amsmath,amsfonts}
\usepackage{algorithmic}
\usepackage{algorithm}
\usepackage{array}

\usepackage{textcomp}
\usepackage{stfloats}
\usepackage{url}
\usepackage{verbatim}
\usepackage{graphicx}
\usepackage{multirow} 
\usepackage{amsmath}  
\usepackage{cite}
\usepackage{amssymb}
\usepackage{subcaption} 
\usepackage{xr}
\usepackage{adjustbox} 
\usepackage[utf8]{inputenc}

\usepackage{xcolor}
\usepackage[colorlinks,linkcolor=black,anchorcolor=black,citecolor=black]{hyperref} 

\graphicspath{{figure/}} 
\def\BibTeX{{\rm B\kern-.05em{\sc i\kern-.025em b}\kern-.08em
T\kern-.1667em\lower.7ex\hbox{E}\kern-.125emX}}
\usepackage{balance}

\begin{document}

\IEEEoverridecommandlockouts

\title{
Energy Minimization Oriented Resource Allocation for Integrated Sensing and Communication in Marine IoT Networks}

\author{Qianru~Wang,~\IEEEmembership{Student~Member,~IEEE,}
	Li~Ping~Qian,~\IEEEmembership{Senior~Member,~IEEE,}
	Chenglong~Dou,
	Haijun~Zhang,~\IEEEmembership{Fellow,~IEEE,}
	and~Yuan~Wu,~\IEEEmembership{Senior~Member,~IEEE}
}

\maketitle

\begin{abstract}
Integrated sensing and communication (ISAC) has become a promising technical framework for Marine Internet of Things (MIoT) systems. Nevertheless, all devices rely on battery power supply, so energy efficiency becomes a core bottleneck that limits practical deployment. This paper thus investigates the energy consumption minimization optimization problem of MIoT-oriented ISAC systems.
In this system, an uncrewed aerial vehicle (UAV) employs non-orthogonal multiple access (NOMA) to simultaneously perform target sensing and collect data from uncrewed surface vehicles (USVs), and subsequently forwards both the processed sensing information and the data collected from the USV to a shore-based base station (SBS). 
Subject to latency limits and sensing performance requirements, the total system energy consumption can be minimized through joint optimization of multiple variables, including the UAV's transmit beamforming, the dedicated sensing signal, the USVs' transmit power, the UAV's computation power, and the time resource allocation for sensing and communication stages. To address the established non-convex optimization problem, we construct a layered solution architecture, which splits the original problem into several separate subproblems and alternately optimizes each subproblem based on its own mathematical characteristics.
In particular, we first derive the closed-form solution for the transmit power of USVs and then apply variable substitution. 
The successive convex approximation (SCA) technique is further adopted to transform the leftover non-convex subproblems into convex forms, based on which we design corresponding efficient iterative solving algorithms. Numerical simulation outcomes verify the validity and accuracy of the proposed algorithm for total system energy reduction. Compared with the orthogonal frequency division multiple access (OFDMA) transmission scheme and the genetic algorithm, our algorithm reduces the system energy consumption by $19.71$\% and $8$\%, respectively. Moreover, the energy consumption optimized by our algorithm only deviates by $8.72$\% from the optimal value solved by the LINGO solver.

\end{abstract}


\section{Introduction}
The fast expansion of the Marine Internet of Things (MIoT) technologies greatly promotes practical scenarios including smart ocean monitoring, autonomous underwater vehicles, and maritime surveillance \cite{9939173,9996964,11142299}. These applications require ultra-reliable and energy-efficient wireless connectivity in harsh oceanic environments, where limited node energy, dynamic topology, and constrained bandwidth pose major challenges to network performance and longevity \cite{9900235,10155303}. Moreover, MIoT systems increasingly leverage integrated sensing and communications (ISAC) to acquire environmental data and support intelligent maritime operations \cite{10812728,10638097,10233771}. 
However, realizing low-latency data transmission and energy-efficient task execution across wide oceanic regions creates a prominent technical barrier, particularly when the system needs to satisfy simultaneous sensing and real-time communication requirements.

Non-orthogonal multiple access (NOMA) serves as a robust technical support to realize massive device connections and spectrum-efficient communications in next-generation wireless networks \cite{10813009}. 
By enabling multiple terminals to reuse identical time-frequency resource units via superposition coding (SC) and signal recovery with successive interference cancellation (SIC), NOMA significantly improves spectral efficiency while supporting differentiated user requirements \cite{9679390,9502643}. This feature makes NOMA particularly suitable for MIoT networks, where densely deployed sensor nodes frequently contend for limited spectrum resources. Notably, ISAC in IoT scenarios including MIoT often faces symbiotic interaction challenges between sensing and communication as well as security risks. These issues need to be addressed with core performance optimization, and integrating ISAC with 6G key technologies is an important direction to tackle these practical problems \cite{11048760}. Moreover, NOMA effectively suppresses co-channel interference in ISAC systems, enabling reliable communication in dense user environments \cite{10129092}. Applying NOMA to marine environments thus offers a promising approach to enhance both energy utilization and sensing accuracy under resource constraints.

Integrating ISAC into MIoT systems offers a transformative solution to the spectral and energy constraints inherent in ocean environments \cite{10552374,10941898}. 
By combining sensing and communication capabilities under one unified architecture, ISAC enables more efficient spectrum utilization while supporting critical marine operations such as real-time ship detection, environmental monitoring, and underwater navigation \cite{10891341, 9945983, 10770127}. This integration is particularly crucial in maritime scenarios, where conventional communication methods often suffer from severe signal attenuation and limited bandwidth. For ISAC-based MIoT networks, non-orthogonal transmission schemes such as NOMA and RSMA adapt well to resource-constrained marine environments while suppressing co-channel interference between sensing and communication signals \cite{10623368}. However, 
the deployment of ISAC brings distinctive technical hurdles, including interference control difficulties and resource allocation balance conflicts between communication quality and sensing precision. Most existing ISAC efforts for MIoT either neglect the strict resource constraints of ocean environments or fail to jointly optimize communication and sensing with non-orthogonal schemes. To address these issues, this work leverages NOMA to jointly optimize the ISAC design, aiming to minimize energy consumption while meeting stringent latency and sensing performance requirements in dynamic marine networks.

With the rapid expansion of marine monitoring and offshore industrial activities, a single-tier data collection framework relying solely on uncrewed surface vehicles (USVs) or static buoys is no longer sufficient to meet the growing demands of MIoT systems. Uncrewed aerial vehicle (UAV)-assisted ISAC networks, where UAVs serve as mobile relays for both data aggregation and real-time environmental sensing, enable more efficient utilization of communication and computing resources while overcoming the limitations of direct long-distance transmissions \cite{10108290,9759502,10516296}. In this paper, we propose a NOMA-enabled ISAC framework for MIoT, where the UAV assists a group of USVs by enabling uplink communication via NOMA while concurrently performing radar sensing toward a target. 
Specifically, there are two phases. In Phase I, the UAV transmits a radar signal to probe the target, while concurrently collecting echo signals and uplink data from USVs via NOMA. 
In Phase II, the UAV relays the fused sensing and communication data to the shore-side base station (SBS). This integrated framework introduces two main challenges. First, in Phase I, the coexistence of sensing and communication results in inter-functional interference. Thus, it is critical to design an effective NOMA mechanism to mitigate such interference, enhancing communication performance without compromising sensing accuracy. 
Second, the allocation of time lengths for Phase I and Phase II brings an inherent trade-off, and reasonable division of phase time can effectively cut down total system energy consumption while meeting the overall latency constraints. The key innovations of this work are summarized below:

\begin{itemize}
	
	\item We propose an energy consumption minimization framework for ISAC in MIoT networks. In Phase I, the UAV simultaneously performs target sensing and collects data from a set of USVs using NOMA. The raw sensing data are processed into meaningful sensing information. In Phase II, the UAV transmits both the USVs' data and sensing information to the SBS.
	To reduce the total energy consumption of UAV and USVs, we jointly optimize the transmit power of USVs, the transmit beamforming and computation power of the UAV, the dedicated sensing signal, and the phase time allocation, under the limits of latency and sensing rate requirements.

	\item 
	Considering the non-convex characteristic of the constructed optimization problem, we take advantage of its layered architecture and split the original model into two separate subproblems belonging to upper and lower layers. We further break down the lower-layer optimization task into three independent sub-tasks. Specifically, we first work out an explicit closed-form formula to calculate the transmit power assigned to each USV. Then, we propose a polyblock approximation-based algorithm to determine the UAV's optimal computation power based on its monotonicity nature. Third, by identifying the rank-one structure of the UAV’s transmit beamforming, we design an alternating optimization-based algorithm to obtain its solution. Finally, we further adopt successive convex approximation (SCA) to optimize the dedicated sensing signal and the durations in these two communication phases, respectively.
	
	\item 
	We carry out abundant numerical simulations to validate the performance and energy-saving capacity of the designed algorithm, and reveal its performance improvements against several mainstream benchmark strategies. The simulation data reveal that total system energy consumption can be cut by an average of $19.71$\% compared with the OFDMA benchmark scheme.

\end{itemize}

The subsequent chapters of this paper are arranged in the following structure. Section \uppercase\expandafter{\romannumeral2} summarizes existing relevant research works. Section \uppercase\expandafter{\romannumeral3} illustrates the system setup and the formulated optimization problem. A high-performance solver built on decomposition theory is put forward in Section \uppercase\expandafter{\romannumeral4} to find the optimal solution. Section \uppercase\expandafter{\romannumeral5} provides numerical simulation data to demonstrate the performance of the designed algorithm. At last, Section \uppercase\expandafter{\romannumeral6} draws the whole paper to a close and puts forward possible directions for subsequent research.

\section{Related Work}
In this section, we first review related studies on NOMA-assisted communication and resource allocation. Then, we review studies on UAV-assisted data collection and ISAC. 

NOMA-assisted communication and resource allocation have gained significant research attention for enhancing system performance. 
In \cite{10376220}, Ting et al. constructed a collaborative optimization architecture targeting computation offloading and resource partitioning within UAV emergency communication systems supported by NOMA in order to cut computational overhead.
In \cite{10946201}, Dai et al. introduced a hybrid semantic communication architecture for maritime networks to achieve energy minimization via joint optimization. 
In \cite{10400811}, Alishahi et al. 
put forward a collaborative resource distribution strategy for IRS-assisted FL-WPT systems integrated with NOMA to cut down system energy consumption.
In \cite{10470406}, Li et al. 
explored mobility-adaptive task offloading and spectrum allocation for NOMA-supported MEC vehicular networks to shorten task processing delay.
In \cite{10839333}, Xu et al. 
developed a unified resource scheduling strategy for NOMA-assisted IoT-MEC networks to reduce weighted cumulative energy consumption.
In \cite{10726693}, Dou et al. devised a joint optimization framework for AP beamforming and two-tier resource allocation in NOMA-assisted ISTTO systems to minimize total energy consumption.
However, most existing works in NOMA-assisted communication focus on general scenarios and fail to consider the specific energy constraints and dynamic environment of MIoT.

UAVs act as a core carrier to deploy integrated sensing and communication systems, and they take advantage of flexible mobile characteristics and the dual-function feature of ISAC to support simultaneous target sensing and high-efficiency communication in complex environments. In \cite{10659350}, Pang et al. 
put forward a UAV-supported sensing-aided transmission strategy for vehicle networking scenarios built upon ISAC, which can dynamically tune sensing and communication performance. In \cite{10410213}, Liu et al. developed a UAV-aided ISAC architecture for IoT to boost radar estimation rate under communication rate constraints. In \cite{10680299}, Ata Khalili et al. 
explored collaborative resource scheduling and flight path planning for multi-terminal multi-target UAV-ISAC frameworks to cut overall power consumption. In \cite{10329470}, Pan et al. focused on a UAV-enabled ISAC system with OFDMA to minimize target location estimation error while ensuring communication QoS. In \cite{10638135}, Gu et al. studied UAV trajectory design for ISAC, addressing trade-offs in different scenarios. In \cite{10680585}, Qin et al. 
built a NOMA-supported UAV ISCC network and adopted deep reinforcement learning to jointly adjust UAV flight trajectory, computing resource allocation, and beamforming vectors.
However, existing UAV-assisted data collection and ISAC works neglect the high dynamics and uncertainty of the marine environment, energy limitations of UAVs, and the stringent latency and real-time requirements of sensing and communication in MIoT.

\begin{table*}[htbp]
	\centering
	\caption{Comparison of Related Works With Our Work}
	\label{tab:comparison}
	\resizebox{\textwidth}{!}{
		\begin{tabular}{|c|c|c|c|c|c|}
			\hline
			\textbf{Access Scheme} & \textbf{Optimization Objective} & \textbf{Reference} & \textbf{Latency Requirement} & \textbf{Computation Resource Allocation} & \textbf{ISAC in UAVs}\\ \hline
			\multirow{5}{*}{OMA} & Achievable Rate Maximization & {\cite{10659350}} & \text{\sffamily X} & \text{\sffamily X} & \checkmark \\ \cline{2-6} 
			&  Radar Estimation Rate Maximization & {\cite{10410213}} & \text{\sffamily X} & \text{\sffamily X} & \checkmark  \\ \cline{2-6} 
			&  Power Consumption Minimization  & {\cite{10680299}} & \text{\sffamily X} & \text{\sffamily X} & \checkmark   \\ \cline{2-6} 
			&  Target Location Estimation Error Minimization    & {\cite{10329470}} & \text{\sffamily X} & \text{\sffamily X} & \checkmark  \\ \cline{2-6} 
			&  Communication Throughput Maximization & {\cite{10638135}} & \text{\sffamily X} & \text{\sffamily X} & \checkmark  \\ \hline
			\multirow{8}{*}{NOMA} & Computation Overhead Minimization  & {\cite{10376220}} & \checkmark & \checkmark & \text{\sffamily X} \\ \cline{2-6} 
			& Energy Consumption Minimization  & {\cite{10946201}} & \checkmark & \text{\sffamily X} & \text{\sffamily X}  \\ \cline{2-6} 
			& Energy Consumption Minimization  & {\cite{10400811}} & \checkmark & \text{\sffamily X} & \text{\sffamily X}   \\ \cline{2-6} 
			& Task Latency Minimization  & {\cite{10470406}} & \checkmark & \checkmark & \text{\sffamily X}   \\ \cline{2-6}
			& Energy Consumption Minimization  & {\cite{10839333}} & \checkmark & \text{\sffamily X} & \text{\sffamily X}  \\ \cline{2-6}
			& Energy Consumption Minimization  & {\cite{10726693}} & \checkmark & \text{\sffamily X} & \checkmark  \\ \cline{2-6} 
			& Computation Throughput Maximization  & {\cite{10680585}} & \text{\sffamily X} & \checkmark & \checkmark  \\ \cline{2-6}
			& Energy Consumption Minimization  & {our work} & \checkmark & \checkmark & \checkmark  \\\hline
		\end{tabular} 
	}
\end{table*}
Table I presents a comprehensive comparison among existing studies and our work across several key dimensions, including access schemes, optimization objectives, latency requirements, computation resource allocation, and ISAC integration in UAV-assisted networks.
Although NOMA-based communication, resource optimization, UAV-assisted data collection, and ISAC have seen progress, challenges remain in MIoT networks. The marine environment's high dynamics and energy limitations of UAVs, along with strict latency requirements, call for optimized transmission and computational resources. 
To tackle these problems, we construct an energy consumption minimization for the ISAC framework in MIoT networks.
We adjust the UAV’s transmit beamforming, the dedicated sensing signal, the transmit power of all USV terminals, the UAV's computation capacity, and phase time allocation together to cut total system energy expenditure and meet latency and sensing performance limits.

\begin{table}[h!] 
	\scriptsize 
	\caption{NOTATIONS AND DEFINITIONS}
	\centering
	\begin{tabular}{ll}
			\hline
			Symbols & Descriptions \\
			\hline
	    	$B^{\text{UAV}}$ & The bandwidth from USV to the UAV \\
			$B^{\text{BS}}$ & The bandwidth from the UAV to the SBS \\
			$\mathbf{c}_n$ & The communication receive beamforming of the UAV \\
			$C^{\text{UAV}}$ & The UAV's required CPU cycles to process each single bit \\ 
			$D_n$ & The volume of data transmitted by USV $n$ \\
			$D^{\text{target}}$ & The transmitted compressed sensing data \\
			$E_n^{\text{USV}}$ & The energy consumption of USV $n$ \\
			$E_{11}^{\text{UAV}}$ & The energy consumption of the UAV in $T_{11}$ \\
			$E^\text{UAV}_{12}$ & The energy consumption of the UAV in $T_{12}$ \\
			$E^{\text{UAV,hov}}$ & The hovering energy consumption of the UAV \\
			$E^{\text{tot}}$ & The total energy consumption of the system \\
			$f^\text{UAV}$ & The UAV's CPU frequency \\
			$f^{\text{UAV},\max}$ & The maximum computing power of the UAV \\
			$\mathbf{g}_n^{t2}$ & The channel gain from USV $n$ to UAV \\
			$\mathbf{H}_{\text{SI}}$ & The residual self-interference channel at the UAV \\
			$\textbf{h}$ & The channel gain between the UAV and the SBS \\
			$\mathbf{n}_0$ & The noise at the UAV \\
			$n_{\text{BS}}$ & The noise at the SBS \\
			$p_{n}$ & The transmit power by USV $n$ to the UAV \\
			$p^{\text{UAV},\max}$ & The maximum transmit power of the UAV \\
			$p_n^{\text{USV},\max}$ & The maximum transmit power of the USV $n$ \\
			$Q$ & The sensing estimation rate of the UAV \\
			$Q^{\text{req}}$ & The required sensing estimation rate \\
			$R_n^{\text{USV}}$ & The achievable transmission rate from USV $n$ to the UAV \\
			$R^{\text{UAV}}$ & The transmission rate from the UAV to the SBS \\
			$\mathbf{s}_0$ & The dedicated sensing signal \\
			$T_1$ & The duration in the first phase \\
			$T_{11}$ & The duration in the phase of collecting the sensed target data\\ 
			$T_{12}$ & The duration in the phase of converting the sensed data \\
			$T_2$ & The duration in the second phase \\
			$T^{\text{tot},\max}$ & The maximum total latency of the system \\
			$T_1^{\max}$ & The maximum latency of the first phase \\
			$T_2^{\max}$ & The maximum latency of the second phase \\
			$\textbf{u}$ & The UAV's sensing receive beamforming \\
			$v$ & The average velocity of the rotor \\
			$\textbf{w}$ & The transmit beamforming for the based station \\
			$w^{\text{UAV}}$ & The weight of the UAV \\
			$\mathbf{x}_{0}$ & The transmitted sensing signal of the UAV \\
			$x_{n}$ & The transmitted signal of USV $n$ \\
			$\widetilde{\textbf{x}}_0$ & The transmitted signal of the UAV in second phase \\
			$\mathbf{y}_0$ & The received signal at the UAV \\
			$\widetilde{\textbf{y}}_{\text{BS}}$ & The received signal at the SBS \\
			$\omega^\text{UAV}$ & The GPU power-efficiency of the UAV \\
			$\delta$ & The duty factor \\
			$\tau$ & The pulse duration \\
			$\boldsymbol \Lambda$ & The total interference of the UAV for sensing the target \\
			$\mathbf{\Gamma}_n$ & The total interference of the UAV for receiving USV $n$'s data \\  
			\hline
		\end{tabular}
	\label{table:1} \vspace{-0.2in}
\end{table}

\begin{figure}[htbp]
	\centering
	\includegraphics[scale=0.26]{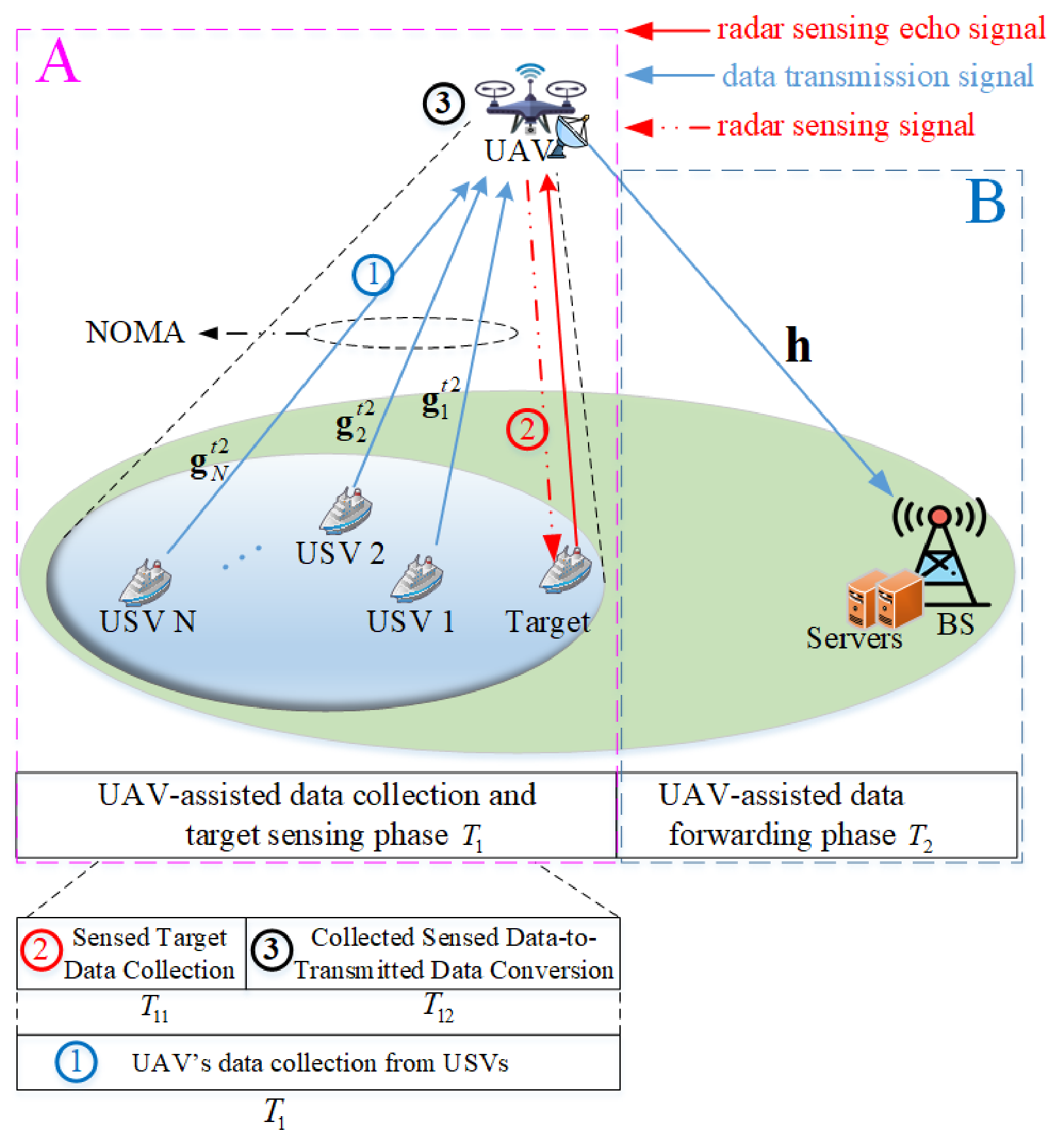} 
	\caption{The MIoT network system model with ISAC. Part A presents the UAV-assisted data collection and target sensing phase, which includes \textcircled{1} the duration of UAV’s data collection from USVs, \textcircled{2} the duration of Sensed Target Data Collection, and \textcircled{3} the duration of Collected Sensed Data-to-Transmitted Data Conversion. \textcircled{1} is carried out simultaneously with \textcircled{2} and \textcircled{3}. Part B presents the UAV-assisted data forwarding phase.}
	\label{fig:model} 
\end{figure}
\section{System model and problem formulation}\label{section:model}
We investigate the problem of energy consumption minimization for ISAC in MIoT networks. 
As shown in Fig. \ref {fig:model}, the MIoT network is composed of multiple single-antenna USVs denoted as $\mathcal{N} = \{1, 2,\ldots ,N\}$, a single-antenna SBS, and one UAV equipped with $N_t$ transmit antennas and $N_r$ receive antennas.
The UAV employs ISAC technology to simultaneously receive data from the USVs and sense a target node. Specifically, it performs radar sensing towards a given target located at angle $\theta_0$, with a required estimation accuracy denoted by $Q^{\text{req}}$. The UAV transmits both the sensed information and the data collected from the USV to the SBS via wireless communication. 
We adopt a 3-D Cartesian coordinate framework to describe the spatial positions of all USVs and the UAV, and fix the vertical coordinate of the sea surface to zero.
Let $q_n = (x_n, y_n, z_n)$ represent the coordinate of USV $n$ with $x_n$ $y_n$ and $z_n$ denoting its x-coordinate, y-coordinate, and z-coordinate respectively. 
The UAV’s coordinate is recorded as $U = (X, Y, H)$ where $X$ and $Y$ stand for horizontal positions and $H$ refers to its steady hovering height. Similarly, we mark the coordinate of the SBS as $S = (X_S, Y_S, Z_S)$ where $X_S$, $Y_S$ and $Z_S$ correspond to its three-dimensional spatial dimensions respectively.
Table \uppercase\expandafter{\romannumeral2} lists the key notations used in this paper.
For clarity, the operational process illustrated in Fig. \ref{fig:model} is split into two independent stages. 
\begin{itemize}
	\item  During the duration in the UAV-assisted data collection and target sensing phase \( T_1 \), the USVs transmit their data to the UAV, while the UAV simultaneously emits a sensing signal toward the target. 
	We adopt the NOMA technique, so the UAV can capture the reflected sensing echo and the transmission signals sent from every USV at the same time. The raw sensing data are subsequently processed into meaningful perception information.
	
	\item  During the duration in the UAV-assisted data forwarding phase \( T_2 \), the UAV transmits both the data collected from the USV and the processed sensing information of the target node to the SBS.
\end{itemize}

\subsection{UAV-assisted Data Collection and Target Sensing Phase}
In this phase, the UAV leverages NOMA to simultaneously collect data $\{D_n\}_{n \in \mathcal{N}}$ from all USVs, where USVs adopt NOMA transmission to send data to the UAV concurrently, a transmission method also employed in  \cite{9996964}, and perform target sensing toward angle $\theta_0$, from which raw sensing data are processed into perception information.

\textit{1) Signal Model:} The transmitted signal of USV $n$ is written as 
\begin{equation}
	\begin{aligned}
		\label{xn}
		x_{n}=\sqrt{p_n} s_n, ~ \forall n \in \mathcal N,
	\end{aligned}
\end{equation}
in which $p_n$ stands for the transmission power used to send the information symbol $s_n$, and this symbol satisfies $\mathbb{E}\{|s_n|^2\}=1$. The transmitted sensing signal of the UAV within the first operational stage is formulated as
\begin{equation}
	\begin{aligned}
		\label{x0}
		\mathbf{x}_{0}=\mathbf{s}_0,
	\end{aligned}
\end{equation}
where $\mathbf{s}_0 \in \mathbb{C}^{N_t \times 1}$ denotes the dedicated sensing signal with covariance $\mathbf{S}_0=\mathbb{E}\{\mathbf{s}_0 (\mathbf{s}_0)^H\}$. 
We establish the sensing channel model linking the UAV and the sensing target. We suppose that the UAV’s transmitting and receiving uniform linear antenna arrays adopt half-wavelength element separation. Accordingly, the transmit steering vector and receive steering vector corresponding to angle $\theta$ take the following forms
\begin{equation}
	\begin{aligned}
		\label{attheta}
		\mathbf{a_t}(\theta)=\frac{1}{\sqrt{N_t}}[1, e^{j \pi \sin \theta}, \ldots ,  e^{j \pi (N_t-1) \sin \theta}]^T,
	\end{aligned}
\end{equation}
\begin{equation}
	\begin{aligned}
		\label{artheta}
		\mathbf{a_r}(\theta)=\frac{1}{\sqrt{N_r}}[1, e^{j \pi \sin \theta}, \ldots ,  e^{j \pi (N_r-1) \sin \theta}]^T.
	\end{aligned}
\end{equation}
We consider that there exist $M$ clutters under the angles of $\{\theta_m\}_{m \in \{1,2,\ldots,M\}}$, and the UAV has the prior knowledge of the target and clutters. 
Given the air-to-sea transmission environment, the transmission link connecting each USV and the UAV, denoted as $\mathbf{g}_n^{t2} \in \mathbb{C}^{N_r \times 1}$, follows Line-of-Sight (LoS) transmission characteristics. Then, the channel coefficient linking USV $n$ to the UAV takes the form $ \mathbf{g}_n^{t2} = \frac{h_0}{\sqrt{\| U - q_n \|^2 + H^2}} \cdot \mathbf{a}_r(\theta_n) $,
where $ h_0 $ stands for the LoS channel gain at the reference distance $ d_0 = 1 \, \text{m} $. 
Therefore, the received signal at the UAV can take the following form 
\begin{equation}
	\begin{aligned}
		\label{y0}
		\mathbf{y}_0 =& \underbrace{\sum_{n=1}^{N} \mathbf{g}_n^{t2}\sqrt{p_n}s_n}_{\text{signal from all USVs}} + \underbrace{\beta_0\mathbf{a_r}(\theta_0)(\mathbf{a_t}(\theta_0))^H\mathbf{x}_0}_{\text{sensing echo from the desired target}} \\
		&+ \underbrace{\sum_{m=1}^{M}\beta_m\mathbf{a_r}(\theta_m)(\mathbf{a_t}(\theta_m))^H\mathbf{x}_0}_{\text{clutter interference}} + \underbrace{\mathbf{H}_{\text{SI}}\mathbf{x}_0 + \mathbf{n}_0}_{\text{self-interference and noise}},
	\end{aligned}
\end{equation}
where 
the complex amplitudes $\{\beta_m\}_{m=0,1,2,\ldots,M}$ are mainly decided by multiple factors including path loss and radar cross-section.
$\mathbf{H}_{\text{SI}} \in \mathbb{C}^{N_r \times N_t}$ stands for the residual self-interference channel coefficient existing on the UAV receiver.
$\mathbf{n}_0$ denotes the noise with variance $\sigma_0^2$. For clarity of notation, we further define $\mathbf{A}_s=\beta_0\mathbf{a_r}(\theta_0)(\mathbf{a_t}(\theta_0))^H$ and $\mathbf{A}_c=\sum_{m=1}^{M}\beta_m\mathbf{a_r}(\theta_m)(\mathbf{a_t}(\theta_m))^H + \mathbf{H}_{\text{SI}}$.

\textit{2) Communication Model:} Within this operational stage, the UAV adopts the NOMA technique to recover incoming signals sent by all USVs.
We denote the decoding order of the USV data streams by a permutation $\Theta$. For instance, $\Theta(m) < \Theta(n)$ indicates that the data from USV $m$ is decoded and subtracted from the received waveform prior to processing the information of USV $n$.
Due to practical hardware constraints, the perfect SIC is hard to achieve, leading to the residual interference from the direct link. According to \cite{9802726, 10623810}, the imperfect SIC effect can be effectively characterized by a linear function model. Thus, the achievable transmission rate $R_n^{\text{USV}}$ for the uplink transmission from USV $n$ to the UAV takes the subsequent form:
\begin{equation}
	\begin{aligned}
		\label{RnUSV}
		R_n^{\text{USV}}=B^{\text{UAV}}\log(1+\frac{p_n \mathbf{c}_n^H \mathbf{g}_n^{t2} (\mathbf{g}_n^{t2})^H \mathbf{c}_n}{\mathbf{c}_n^H \mathbf{\Gamma}_n \mathbf{c}_n}), ~ \forall n \in \mathcal N,
	\end{aligned}
\end{equation}
where $B^{\text{UAV}}$ denotes the transmission channel bandwidth in this phase, $\mathbf{c}_n \in \mathbb{C}^{N_r \times 1}$ represents the receiving beamforming vector deployed on the UAV for communication,
and $\mathbf{\Gamma}_n$ refers to the aggregate interference encountered while decoding the data stream of USV $n$, and this interference term takes the following form
\begin{equation}
	\small
	\begin{aligned}
		\label{Gamman}
		\mathbf{\Gamma}_n=&\sum_{\Theta(m) > \Theta(n)}p_m \mathbf{g}_m^{t2} (\mathbf{g}_m^{t2})^H + \mathbf{A}_s\mathbf{S}_0\mathbf{A}_s^H + \mathbf{A}_c\mathbf{S}_0\mathbf{A}_c^H \\
		&+ \sigma_0^2\mathbf{I}_N + \zeta \sum_{\Theta(m) < \Theta(n)}p_m \mathbf{g}_m^{t2} (\mathbf{g}_m^{t2})^H, ~ \forall n \in \mathcal N.
	\end{aligned}
\end{equation}
where $\zeta$ represents the SIC factor that quantifies the residual decoded interference in NOMA.  Specifically, $\zeta = 1$ means that no SIC is conducted, while $\zeta = 0$ denotes the
perfect SIC scenario. The specific value of $\zeta$ depends on hardware implementation. Considering the research focus of this paper, $\zeta$ is set to 0 in the following derivations.

According to \cite{10726693}, the optimal receiving vector $\mathbf{c}_n^\ast$ takes the following form
\begin{equation}
	\small
	\begin{aligned}
		\label{cnast}
		\mathbf{c}_n^\ast=\text{arg}\max \frac{p_n \mathbf{c}_n^H \mathbf{g}_n^{t2} (\mathbf{g}_n^{t2})^H \mathbf{c}_n}{\mathbf{c}_n^H \mathbf{\Gamma}_n \mathbf{c}_n} = \mathbf{\Gamma}_n^{-1}\mathbf{g}_n, ~ \forall n \in \mathcal N .
	\end{aligned}
\end{equation}
We plug the solved optimal vector $\mathbf{c}_n^\ast$ into Eq. \eqref{RnUSV}, so we can restate the attainable uplink data rate in a new expression
\begin{equation}
	\begin{aligned}
		\label{RnUSV2}
		R_n^{\text{USV}}=B^{\text{UAV}}\log(1+p_n (\mathbf{g}_n^{t2})^H \mathbf{\Gamma}_n^{-1} \mathbf{g}_n^{t2}), ~ \forall n \in \mathcal N.
	\end{aligned}
\end{equation}
With the duration $T_1$, the achievable transmission rate should satisfy
\begin{equation}
	\begin{aligned}
		\label{RnUSV3}
		T_1 R_n^{\text{USV}}=D_n, ~ \forall n \in \mathcal N.
	\end{aligned}
\end{equation}
The total energy consumption generated by USV $n$ takes the form
\begin{equation}
	\begin{aligned}
		\label{EnUSV}
		E_n^{\text{USV}}=p_n T_1, ~ \forall n \in \mathcal N.
	\end{aligned}
\end{equation}

\textit{3) Sensing Model:} 
To handle the sensing module, we divide the duration $T_1$ into two parts show in Fig. \ref{fig:model}, the duration time $T_{11}$ for collecting the sensed target data and the duration time $T_{12}$ for converting the collected sensed data into transmitted data, so the duration $T_1$ can be expressed as
\begin{equation}
	\begin{aligned}
		\label{T1=T11+T12}
		T_1 = T_{11}+T_{12}.
	\end{aligned}
\end{equation}
We utilize the sensing estimation rate\footnote{The sensing estimation rate measures the reduction rate of target parameter uncertainty per second. It is suitable for the real-time ISAC systems with latency constraints. Unlike the CRB, which gives a lower bound of estimation error variance for static parameter estimation, the sensing estimation rate emphasizes dynamic sensing efficiency, which aligns better with our marine IoT scenario.} \cite{10726693} to measure the radar sensing performance, which characterizes the elimination degree of target parameter uncertainties within each unit time.
Based on Eq. \eqref{y0}, after decoding the transmission signal from all USVs, the sensing estimation rate $Q$ of the UAV for sensing the target in this phase can be expressed as
\begin{equation}
	\begin{aligned}
		\label{Q}
		Q=\frac{\delta}{2\tau}\log(1+\frac{2 \tau B^{\text{UAV}} \textbf{u}^H \textbf{A}_s \textbf{S}_0 (\textbf{A}_s)^H \textbf{u}}{\textbf{u}^H \boldsymbol \Lambda \textbf{u}})
	\end{aligned}
\end{equation}
in which $\delta$ stands for the duty cycle, while $\tau$ refers to single pulse time length.
The vector $\textbf{u}$ denotes the UAV's sensing receive beamforming. In Eq. \eqref{Q}, $\boldsymbol \Lambda$ denotes the total interference of the UAV for sensing the target in this phase, and this interference term is formulated as
\begin{equation}
	\begin{aligned}
		\label{Lambda}
		\boldsymbol \Lambda=\textbf{A}_c \textbf{S}_0 \textbf{A}_c^H + \sigma_0^2\mathbf{I}_N.
	\end{aligned}
\end{equation}
Deriving from the research outcome of reference \cite{10726693}, the optimal receiving beamforming vector $\mathbf{u}^\ast$ takes the following form
\begin{equation}
	\begin{aligned}
		\label{uast}
		\mathbf{u}^\ast=\text{arg}\max \frac{\textbf{u}^H \textbf{A}_s \textbf{S}_0 (\textbf{A}_s)^H \textbf{u}}{\textbf{u}^H \boldsymbol \Lambda \textbf{u}} = \boldsymbol{\Lambda}^{-1} \mathbf{a_r}(\theta_0).
	\end{aligned}
\end{equation}
We substitute the solved optimal vector $\mathbf{u}^\ast$ into Eq. \eqref{Q}, which allows us to restate the target sensing estimation rate as
\begin{equation}
	\begin{aligned}
		\label{Q2}
		Q=\frac{\delta}{2\tau}\log(1+ 2 \tau B^{\text{UAV}} (\textbf{s}_0)^H \textbf{A}_s^H \boldsymbol{\Lambda}^{-1} \textbf{A}_s \textbf{s}_0).
	\end{aligned}
\end{equation}
With the duration $T_{11}$, the energy consumption produced by the UAV platform is formulated in the below expression:
\begin{equation}
	\begin{aligned}
		\label{E0UAV}
		E_{11}^{\text{UAV}}=\text{Tr}(\textbf{S}_0) T_{11}.
	\end{aligned}
\end{equation}

\textit{4) Computational model:} After receiving the sensing information, the UAV processes and converts it into transmission-ready data $D^{\text{target}}$ through techniques such as feature extraction, compression, and encoding. It can be expressed as
\begin{equation}
	\begin{aligned}
		\label{D^target}
		D^{\text{target}}=\alpha Q T_{11},
	\end{aligned}
\end{equation}
where \( \alpha \in (0, 1] \) denotes the compression efficiency coefficient \cite{11005494}, a dimensionless parameter that quantifies the relationship between the volume of sensing data received by the UAV and the sensing data transmitted by the UAV.
The duration $T_{12}$ can be expressed as
\begin{equation}
	\begin{aligned}
		\label{T12}
		T_{12} = \frac{Q T_{11} C^{\text{UAV}}}{f^\text{UAV}},
	\end{aligned}
\end{equation}
in which $f^\text{UAV}$ corresponds to the CPU operating frequency of the UAV, 
and $C^{\text{UAV}}$ denotes the UAV's required CPU cycles for each bit.

With the duration $T_{12}$, the total energy consumption generated by the UAV platform is expressed in the below form:
\begin{equation}
	\begin{aligned}
		\label{E12}
		E^\text{UAV}_{12} = T_{12} \omega^\text{UAV} (f^\text{UAV})^3 = \omega^\text{UAV} Q T_{11} C^{\text{UAV}} (f^\text{UAV})^2,
	\end{aligned}
\end{equation}
where $\omega^\text{UAV}$ is the GPU power-efficiency of the UAV.

\subsection{UAV-assisted Data forwarding Phase}
After collecting all data from the USVs and processing the sensing data, the UAV forwards the results to the SBS.

\textit{1) Signal Model:} The transmitted signal by the UAV within this operational stage takes the form
\begin{equation}
	\begin{aligned}
		\label{x0T}
		\widetilde{\textbf{x}}_0=\textbf{w}z,
	\end{aligned}
\end{equation}
in which $\textbf{w} \in \mathbb{C}^{N_t \times 1}$ stands for the transmit beamforming vector dedicated to the SBS, and $z$ satisfying $\mathbb{E}\{|z|^2\}=1$ corresponds to the data symbol intended for the SBS. We assume the communication link connecting the UAV and SBS follows line-of-sight transmission, and its channel coefficient takes the form $ \mathbf{h} = \frac{h_0}{\sqrt{\| U - S \|^2 + H^2}} \cdot \mathbf{a}_t(\theta) $. The signal captured by the SBS receiving terminal adopts the following expression
\begin{equation}
	\begin{aligned}
		\label{yBST}
		\widetilde{\textbf{y}}_{\text{BS}}=\textbf{h}^H \textbf{w} z + n_{\text{BS}},
	\end{aligned}
\end{equation}
in which $\textbf{h} \in \mathbb{C}^{N_t \times 1}$ stands for the channel coefficient over the air link between the UAV platform and ground SBS, while $n_{\text{BS}}$ refers to the additive Gaussian noise with variance $\sigma_{\text{BS}}^2$.

\textit{2) Communication Model:} In this phase, the transmission rate $R^{\text{UAV}}$ achieved by the SBS can be expressed as
\begin{equation}
	\begin{aligned}
		\label{RUAV}
		R^{\text{UAV}}=B^{\text{BS}}\log(1+ \frac{|\textbf{h}^H \textbf{w}|^2}{\sigma_{\text{BS}}^2}),
	\end{aligned}
\end{equation}
where $B^{\text{BS}}$ denotes the transmission channel bandwidth between the UAV and the SBS. With the duration $T_2$, the transmission data should satisfy
\begin{equation}
	\begin{aligned}
		\label{Dtot}
		\sum_{n=1}^{N}D_n + D^{\text{target}} \le R^{\text{UAV}} T_2.
	\end{aligned}
\end{equation}
During this phase, the total energy consumption generated by the UAV platform is formulated in the below expression:
\begin{equation}
	\begin{aligned}
		\label{EUAV}
		E^{\text{UAV}}=\text{Tr}(\textbf{w} \textbf{w}^H) T_2.
	\end{aligned}
\end{equation}

\subsection{Total Energy Consumption}
Before calculating the overall energy consumption of the system, we need to take the hovering energy consumption of the UAV within the two time segments into consideration, and this energy term takes the following form
	\begin{equation}
		\begin{aligned}
			\label{EUAVhov}
			E^{\text{UAV,hov}}=p^{hov} (T_1 + T_2),
		\end{aligned}
	\end{equation}
in which $p^{hov}$ stands for the steady power consumed to maintain the UAV’s hovering state, and this parameter is mainly determined by multiple factors including the aircraft weight, air density, rotor efficiency and lift magnitude. 
According to \cite{9080561,9411725}, it can be expressed as 
	\begin{equation}
		\begin{aligned}
			\label{phov}
			p^{hov}=\frac{{w^{\text{UAV}}}^{3/2}}{\sqrt{2\rho A C_T}},
		\end{aligned}
	\end{equation}
	where $w^{\text{UAV}}$ denotes the weight of the UAV, $\rho$ refers to the ambient air density, and $A$ represents the rotor disc area, which is typically expressed as the square of the rotor radius $r$ multiplied by $\pi$, i.e., $A = \pi r^2$, and $C_T$ denotes the thrust coefficient. Notably, most existing works adopt the ideal momentum theory model with \( C_T = 1 \) to simplify UAV hovering energy consumption by neglecting the thrust coefficient \( C_T \). However, practical rotors exhibit aerodynamic losses such as blade element drag and uneven airflow. Following \cite{https://doi.org/10.1155/2018/9632942}, we introduce \( C_T \) to refine the ideal model, thereby enhancing the accuracy of hovering power characterization. It can generally be expressed as
	\begin{equation}
		\begin{aligned}
			\label{CT}
			C_T=\frac{L}{\frac{1}{2}\rho A v^2},
		\end{aligned}
	\end{equation}
	where $L$ denotes the lift force provided by the rotor, which is typically equal to the weight of the UAV $w^{\text{UAV}}$ in a hovering state, and $v$ denotes the average velocity of the rotor. According to Eqs. (26), (27), and (28), the hovering energy consumption of the UAV can be rewritten as
	\begin{equation}
		\begin{aligned}
			\label{EUAVhov2}
			E^{\text{UAV,hov}}=\frac{1}{2}v w^{\text{UAV}} (T_1 + T_2).
		\end{aligned}
	\end{equation}
The overall energy consumption generated by the whole system is
\begin{equation}
	\begin{aligned}
		\label{Etot}
		E^{\text{tot}}=\sum_{n=1}^{N}E_n^{\text{USV}}+E_{11}^{\text{UAV}}+E_{12}^{\text{UAV}}+E^{\text{UAV}}+E^{\text{UAV,hov}}.
	\end{aligned}
\end{equation}

\subsection{Problem Formulation}
This work targets the minimization of 
the system energy consumption while guaranteeing both the latency requirement and the sensing rate requirement. 
To reach this optimization objective, we jointly optimize the transmit beamforming of the UAV $\textbf{w}$, the dedicated sensing signal $\textbf{s}_0$, the transmit powers of the USVs $\{p_n\}_{n \in \mathcal{N}}$, the computation power of the UAV $f^\text{UAV}$, and  the durations $T_{11}$ and $T_2$. The formulated optimization problem is presented below:
\begin{subequations} \label{p1}%
	\begin{align}
		\textbf{$\mathbf{P1}$:} ~ & \min  ~ E^\text{tot} \nonumber \\
		&\text{s.t.} ~\text{constraints}~\eqref{RnUSV3},~\eqref{Dtot},\nonumber \\ 
		&\quad ~~ T_{11} + T_{12} + T_2 \le T^{\text{tot},\max} ,      \label{CTtot}\\
		&\quad ~~ Q \ge Q^{\text{req}},     \label{CQ}\\
		&\quad ~~ 0 \le T_1 \le T_1^{\max},      \label{CT1}\\
		&\quad ~~ 0 \le T_2 \le T_2^{\max},      \label{CT2}\\
		&\quad ~~ \text{Tr}(\textbf{w}\textbf{w}^H) \le p^{\text{UAV},\max},      \label{Cw}\\
		&\quad ~~ \text{Tr}(\textbf{S}_0) \le p^{\text{UAV},\max},      \label{CS0}\\
		&\quad ~~ p_n \le p_n^{\text{USV},\max},      \label{Cpn}\\
		&\quad ~~ f^{\text{UAV},\min} \le f^\text{UAV} \le f^{\text{UAV},\max},      \label{Cf}\\
		&\text{vars.}~ \textbf{s}_0,~\textbf{w},~\{p_n\}_{n \in \mathcal{N}},~T_{11},~f^\text{UAV},~T_2.  \nonumber 
	\end{align}%
\end{subequations}
Constraint \eqref{CTtot} ensures that the total latency doesn't exceed the maximum allowable latency $T^{\text{tot},\max}$.
Constraint \eqref{CQ} guarantees the required sensing estimation rate $Q^{\text{req}}$.
Constraints \eqref{CT1} and \eqref{CT2} mean that both the duration in Phase-I $T_1$ and the duration in Phase-II $T_2$ cannot exceed a specified upper-bound.
Constraints \eqref{Cw} and \eqref{CS0} cap the total transmission power radiated by the UAV within its hardware power limit $p^{\text{UAV},\max}$.
Constraint \eqref{Cpn} provides an upper bound for USV's transmit power.
Constraint \eqref{Cf} implies that the UAV's processing-rate cannot exceed its computing capacity $f^{\text{UAV},\max}$.

\section{Algorithm Design} \label{Algorithm Design}
\begin{figure}[t]
	\centering
	\includegraphics[scale=0.17]{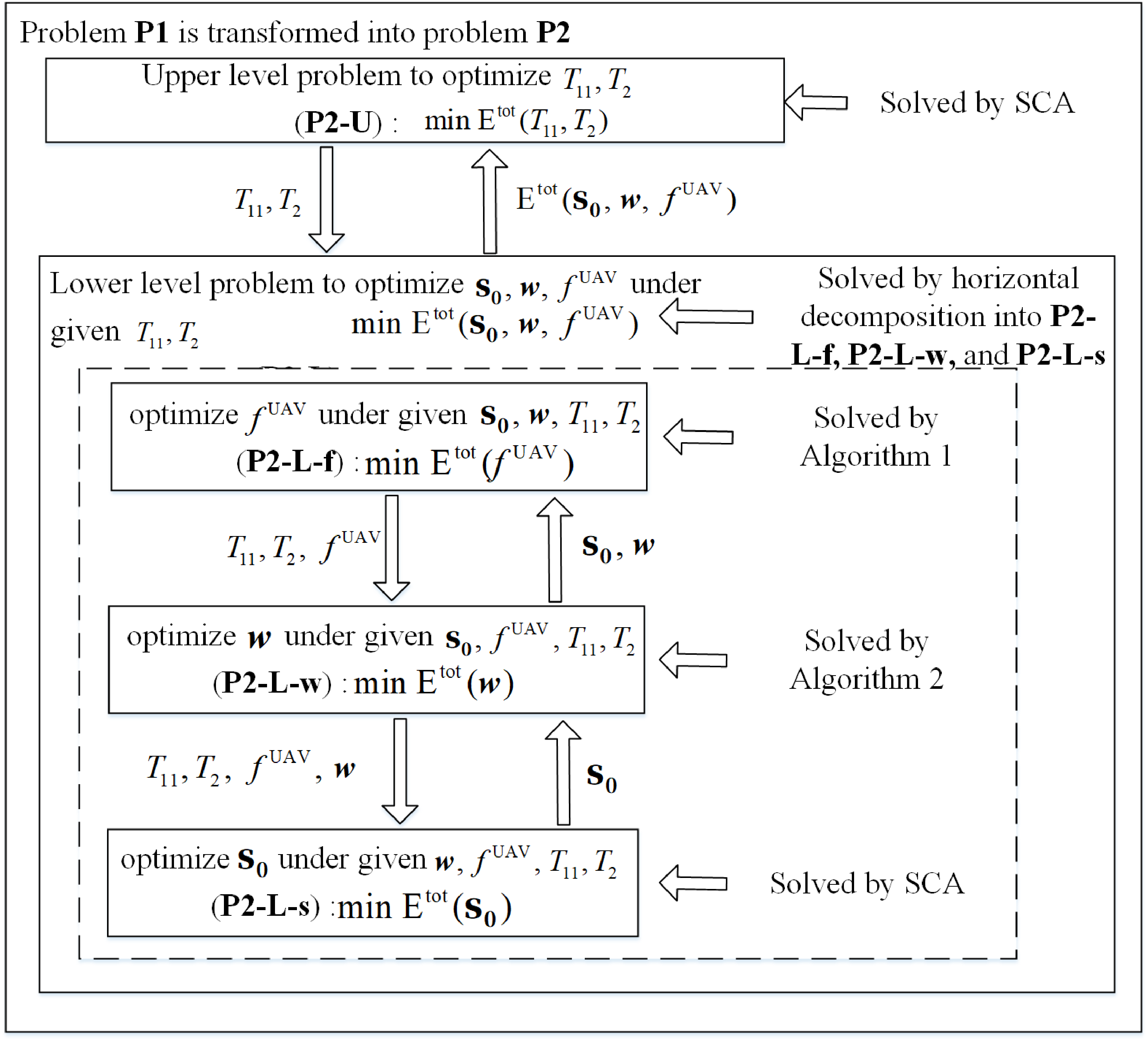} 
	\caption{Decomposition Structure for Solving Problem $\mathbf{P1}$.}
	\label{decomposeP1} 
\end{figure}

In this section, we tackle the optimization problem $\mathbf{P1}$ via the decomposition framework. In detail, we first derive closed-form solutions for all USV transmit power variables and convert the 
problem $\mathbf{P1}$ into an equivalent problem $\mathbf{P2}$. 
Next, we split $\mathbf{P2}$ into two nested subproblems through vertical decomposition: the lower-layer subproblem $\mathbf{P2}\mbox{-}\mathbf{L}$,
which optimizes the UAV’s transmit beamforming, the dedicated sensing signal, and computation power, and a upper level problem $\mathbf{P2}\mbox{-}\mathbf{U}$, which optimizes the durations in the two phases. To solve $\mathbf{P2}\mbox{-}\mathbf{L}$, we apply the block coordinate descent (BCD) method \cite{10726693,11141791,10375321} and further decompose it horizontally into 
three separate subproblems: $\mathbf{P2}\mbox{-}\mathbf{L}\mbox{-}\mathbf{f}$ for tuning the UAV’s computation power, $\mathbf{P2}\mbox{-}\mathbf{L}\mbox{-}\mathbf{w}$ for designing the UAV transmit beamforming and $\mathbf{P2}\mbox{-}\mathbf{L}\mbox{-}\mathbf{s}$ for optimizing the dedicated sensing signal. We put forward an algorithm built on polyblock approximation to handle the optimization subproblem $\mathbf{P2}\mbox{-}\mathbf{L}\mbox{-}\mathbf{f}$.
For problem $\mathbf{P2}\mbox{-}\mathbf{L}\mbox{-}\mathbf{w}$, we leverage the inherent rank-1 characteristic of the transmit beamforming and construct an alternating optimization algorithm for its solution.
For $\mathbf{P2}\mbox{-}\mathbf{L}\mbox{-}\mathbf{s}$ and $\mathbf{P2}\mbox{-}\mathbf{U}$, we apply the SCA method and solve them iteratively using the CVX solver. 
This multi-layer solution framework enables us to efficiently resolve the original optimization P1 through iterative iterations, as illustrated in Fig. \ref{decomposeP1}.

\subsection{Problem Transformation}
We can observe that the optimization formulation $\mathbf{P1}$ falls into the category of strictly non-convex optimization problems. This inherent property brings obstacles to directly solving the original formulation $\mathbf{P1}$. To address this difficulty, we first restructure the mathematical expression of $\mathbf{P1}$ to derive its equivalent optimization form.
 
By putting Eq. \eqref{RnUSV2} into constraint \eqref{RnUSV3}, we can derive
\begin{equation}
	\begin{aligned}
		\label{pn}
		p_n=&\big((\mathbf{g}_n^{t2})^H (\mathbf{A}_s\mathbf{S}_0\mathbf{A}_s^H + \mathbf{A}_c\mathbf{S}_0\mathbf{A}_c^H 
		+ \sigma_0^2\mathbf{I}_N)^{-1} \mathbf{g}_n^{t2}\big)^{-1} \\
		&(2^{\frac{D_n}{(T_{11}+\frac{Q T_{11} C^{\text{UAV}}}{f^\text{UAV}})B^{\text{UAV}}}}-1) 2^{\frac{ \sum_{\Theta(m) > \Theta(n)} D_m}{(T_{11}+\frac{Q T_{11} C^{\text{UAV}}}{f^\text{UAV}})B^{\text{UAV}}}}.
	\end{aligned}
\end{equation}
We rewrite $E^{\text{tot}}$ as 
\begin{equation}
	\begin{aligned}
		\label{Etot2}
		E^{\text{tot}}=&\sum_{n=1}^{N} \big((\mathbf{g}_n^{t2})^H \Omega^{-1} \mathbf{g}_n^{t2}\big)^{-1}  (2^{\frac{D_n}{(T_{11}+\frac{Q T_{11} C^{\text{UAV}}}{f^\text{UAV}})B^{\text{UAV}}}}-1) \\
		& 2^{\frac{ \sum_{\Theta(m) > \Theta(n)} D_m}{(T_{11}+\frac{Q T_{11} C^{\text{UAV}}}{f^\text{UAV}})B^{\text{UAV}}}} (T_{11}+\frac{Q T_{11} C^{\text{UAV}}}{f^\text{UAV}}) \\
		&+\text{Tr}(\textbf{S}_0) T_{11} +\omega^\text{UAV} Q T_{11} C^{\text{UAV}} (f^\text{UAV})^2 +\text{Tr}(\textbf{w} \textbf{w}^H) T_2 \\
		&+\frac{1}{2}v w^{\text{UAV}} (T_{11}+\frac{ Q T_{11} C^{\text{UAV}}}{f^\text{UAV}} +T_2),
	\end{aligned}
\end{equation}
where $\Omega=\mathbf{A}_s\mathbf{S}_0\mathbf{A}_s^H + \mathbf{A}_c\mathbf{S}_0\mathbf{A}_c^H + \sigma_0^2\mathbf{I}_N$.

Then, we can rewrite the original optimization formulation $\mathbf{P1}$ into its equivalent counterpart labeled $\mathbf{P2}$, which is presented below:
\begin{subequations} \label{p2}%
	\begin{align}
		\textbf{$\mathbf{P2}$:} ~ & \min  ~ E^\text{tot} \nonumber \\
		&\text{s.t.} ~\text{constraints}~\eqref{Dtot},~\eqref{CTtot},~\eqref{CQ},~\eqref{CT1},~\eqref{CT2},~\eqref{Cw},\nonumber \\
		&\quad\quad  \eqref{CS0},~\eqref{Cf},\nonumber \\ 
		&\quad ~~ \big((\mathbf{g}_n^{t2})^H \Omega^{-1} \mathbf{g}_n^{t2}\big)^{-1} (2^{\frac{D_n}{(T_{11}+\frac{Q T_{11} C^{\text{UAV}}}{f^\text{UAV}})B^{\text{UAV}}}}-1) \nonumber\\
		&\quad~~ 2^{\frac{ \sum_{\Theta(m) > \Theta(n)} D_m}{(T_{11}+\frac{Q T_{11} C^{\text{UAV}}}{f^\text{UAV}})B^{\text{UAV}}}} \le p_n^{\text{USV},\max},      \label{Cpn2}\\
		&\text{vars.}~ \textbf{s}_0,~\textbf{w},~T_{11},~f^\text{UAV},~T_2.  \nonumber 
	\end{align}%
\end{subequations}

\subsection{Decomposition of Problem $\mathbf{P2}$}
The reformulated optimization formulation $\mathbf{P2}$ still falls into the category of strictly non-convex optimization problem. To address this difficulty, we put forward a layered decomposition strategy as the core solution.

\textit{1) Lower level problem under the given $T_{11}$ and $T_2$:}  We firstly consider that the durations $T_{11}$ and $T_2$ are given, and aim at optimizing the transmit beamforming of the UAV $\textbf{w}$, the dedicated sensing signal $\textbf{s}_0$, and the computation power of the UAV $f^\text{UAV}$. This leads to problem $\mathbf{P2}$ turns into problem $\mathbf{P2}\mbox{-}\mathbf{L}$ as follows:
\begin{subequations} \label{p2-B}%
	\begin{align}
		\textbf{$\mathbf{P2}\mbox{-}\mathbf{L}$:}& ~  \min  ~ E^\text{tot} \nonumber \\
		&\text{s.t.} ~\text{constraints}~\eqref{Dtot},~\eqref{CTtot},~\eqref{CQ},~\eqref{CT1},~\eqref{Cw},\nonumber \\
		&\quad\quad  \eqref{CS0},~\eqref{Cf},~\eqref{Cpn2} \nonumber \\ 
		&\text{vars.}~ \textbf{s}_0,~\textbf{w},~f^\text{UAV}.  \nonumber 
	\end{align}%
\end{subequations}

\textit{2) Upper level problem to optimize $T_{11}$ and $T_2$:} After solving problem $\mathbf{P2}\mbox{-}\mathbf{L}$ and obtaining the value of $E^\text{tot}$ with the given $T_{11}$ and $T_2$, we then continue to optimize $T_{11}$ and $T_2$, which results in the following optimization problem.
\begin{subequations} \label{p2-T}%
	\begin{align}
		\textbf{$\mathbf{P2}\mbox{-}\mathbf{U}$:} ~ & \min  ~ E^\text{tot} \nonumber \\
		&\text{s.t.} ~\text{constraints}~\eqref{Dtot},~\eqref{CTtot},~\eqref{CT1},~\eqref{CT2},~\eqref{Cpn2},\nonumber \\ 
		&\text{vars.}~ T_{11},~T_2.  \nonumber 
	\end{align}%
\end{subequations}

\subsection{Decomposition of Problem $\mathbf{P2}\mbox{-}\mathbf{L}$}
In problem $\mathbf{P2}\mbox{-}\mathbf{L}$, the UAV's CPU frequency $f^\text{UAV}$, the UAV's transmit beamforming $\textbf{w}$, and the dedicated sensing signals $\textbf{s}_0$ show a strong coupling effect within both the objective term and all constraint conditions. To acquire an efficient solution for subproblem $\mathbf{P2}\mbox{-}\mathbf{L}$, 
we propose the following decomposition of problem $\mathbf{P2}\mbox{-}\mathbf{L}$ based on the BCD method.

\textit{1) Sub-problem to optimize $f^\text{UAV}$ under the given $\textbf{w}$ and $\textbf{s}_0$:}
\begin{subequations} \label{p2-B-f}%
	\begin{align}
		\textbf{$\mathbf{P2}\mbox{-}\mathbf{L}\mbox{-}\mathbf{f}$:}& ~  \min  ~ E^\text{tot} \nonumber \\
		&\text{s.t.} ~\text{constraints}~\eqref{CTtot},~\eqref{CT1},~\eqref{Cf},~\eqref{Cpn2}, \nonumber \\ 
		&\text{var.}~ f^\text{UAV}.  \nonumber 
	\end{align}%
\end{subequations}

\textit{2) Sub-problem to optimize $\textbf{w}$ under the given $f^\text{UAV}$ and $\textbf{s}_0$:}
\begin{subequations} \label{p2-B-w}%
	\begin{align}
		\textbf{$\mathbf{P2}\mbox{-}\mathbf{L}\mbox{-}\mathbf{w}$:}& ~  \min  ~ E^\text{tot} \nonumber \\
		&\text{s.t.} ~\text{constraints}~\eqref{Dtot},~\eqref{Cw},\nonumber \\
		&\text{var.}~ \textbf{w}.  \nonumber 
	\end{align}%
\end{subequations}

\textit{3) Sub-problem to optimize $\textbf{s}_0$ under the given $f^\text{UAV}$ and $\textbf{w}$:}
\begin{subequations} \label{p2-B-s}%
	\begin{align}
		\textbf{$\mathbf{P2}\mbox{-}\mathbf{L}\mbox{-}\mathbf{s}$:}& ~  \min  ~ E^\text{tot} \nonumber \\
		&\text{s.t.} ~\text{constraints}~\eqref{Dtot},~\eqref{CTtot},  \nonumber \\
		&~~~~~~~\eqref{CQ},~\eqref{CT1},~\eqref{CS0},~\eqref{Cpn2}, \nonumber \\ 
		&\text{var.}~ \textbf{s}_0.  \nonumber 
	\end{align}%
\end{subequations}

According to the criterion of the BCD method, the three sub-problems (i.e., problem $\mathbf{P2}\mbox{-}\mathbf{L}\mbox{-}\mathbf{f}$, problem $\mathbf{P2}\mbox{-}\mathbf{L}\mbox{-}\mathbf{w}$ and problem $\mathbf{P2}\mbox{-}\mathbf{L}\mbox{-}\mathbf{s}$) are solved iteratively to acquire the solution of the lower-layer formulation $\mathbf{P2}\mbox{-}\mathbf{L}$.
This splitting scheme possesses a primary advantage in that all decomposed subproblems can be handled separately to attain their respective optimal points.

\subsection{Proposed Algorithm for Solving Problem $\mathbf{P2}\mbox{-}\mathbf{L}\mbox{-}\mathbf{f}$}
We introduce the auxiliary variable $x=T_{11}+\frac{Q T_{11} C^{\text{UAV}}}{f^\text{UAV}}$, and problem $\mathbf{P2}\mbox{-}\mathbf{L}\mbox{-}\mathbf{f}$ can then be transformed into
\setcounter{equation}{34}
\begin{subequations} \label{p2-B-f'}%
	\begin{align}
		\textbf{$\mathbf{L}\mbox{-}\mathbf{f1}$:}&   \min \sum_{n=1}^{N} \big((\mathbf{g}_n^{t2})^H \Omega^{-1} \mathbf{g}_n^{t2}\big)^{-1} (2^{\frac{D_n}{x B^{\text{UAV}}}}-1) \nonumber \\
		&~~~~ 2^{\frac{ \sum_{\Theta(m) > \Theta(n)} D_m}{x B^{\text{UAV}}}} x \nonumber\\
		&~~~~+\text{Tr}(\textbf{S}_0) T_{11} +\omega^\text{UAV} Q T_{11} C^{\text{UAV}} (\frac{Q T_{11} C^{\text{UAV}}}{x - T_{11}})^2  \nonumber\\
		&~~~~+\text{Tr}(\textbf{w} \textbf{w}^H) T_2 +\frac{1}{2}v w^{\text{UAV}} (x +T_2) \nonumber \\
		&\text{s.t.} 
		~  x \le T^{\text{tot},\max} - T_2,      \label{CTtot-bf'}\\
		&\quad ~~ x \ge \frac{Q T_{11} C^{\text{UAV}}}{f^{\text{UAV},\max}} + T_{11},      \label{Cf-bf'}\\
		&\quad ~~ \big((\mathbf{g}_n^{t2})^H \Omega^{-1} \mathbf{g}_n^{t2}\big)^{-1} (2^{\frac{D_n}{x B^{\text{UAV}}}}-1) 2^{\frac{ \sum_{\Theta(m) > \Theta(n)} D_m}{x B^{\text{UAV}}}} \nonumber\\
		&\quad ~~ \le p_n^{\text{USV},\max},      \label{Cpn2-bf'}\\
		&\quad ~~ 0 \le x \le T_1^{\max},      \label{CT1-bf'}\\
		&\text{var.}~ x.  \nonumber 
	\end{align}%
\end{subequations}
We analyze the monotonic characteristic of the objective function belonging to subproblem 
$\mathbf{L}\mbox{-}\mathbf{f1}$ with respect to 
$x$, as shown in Proposition 1. Full derivations to verify Proposition 1 are provided within Appendix A.

\noindent\textbf{Proposition 1.} The objective function of problem $\mathbf{L}\mbox{-}\mathbf{f1}$ is the difference between two monotonic functions with respect to $x$.

With the implicit monotonic property of subproblem $\mathbf{L}\mbox{-}\mathbf{f1}$ revealed in Proposition 1, 
we can establish problem $\mathbf{L}\mbox{-}\mathbf{f1}$ as a canonical monotonic optimization form by introducing an auxiliary variable $e$ and a new function $H(x,e)$, which are shown as follows:
\begin{equation}
	\begin{aligned} \label{e}
		e=E^{\max}-(E^{\text{UAV,hov}}+E_{11}^{\text{UAV}}+E^{\text{UAV}}),
	\end{aligned}
\end{equation}
\begin{equation}
	\begin{aligned} \label{H(x,e)}
		H(x,e)=e-(E_{12}^{\text{UAV}}+\sum_{n=1}^{N}E_n^{\text{USV}}).
	\end{aligned}
\end{equation}
Here, $E^{\max}$ can be acquired when variable $x$ is assigned its upper bound $x^{\text{max}}$, and the subproblem $\mathbf{L}\mbox{-}\mathbf{f1}$ can be rewritten into its equivalent formulation presented below:
\begin{subequations} \label{max-H(x,e)}%
	\begin{align}
		\textbf{$\mathbf{L}\mbox{-}\mathbf{f2}$:}&   \max H(x,e) \nonumber\\
		&\text{s.t.} ~\text{constraints}~\eqref{Cpn2-bf'} \nonumber \\ 
		&\quad ~~ \max\{0,\frac{Q T_{11} C^{\text{UAV}}}{f^{\text{UAV},\max}} + T_{11}\} \le x \nonumber\\
		&\quad ~~ \le \min \{T_1^{\max},T^{\text{tot},\max} - T_2\},      \label{C-x}\\
		&\quad ~~ 0 \le e \le E^{\max}-(E^{\text{UAV,hov}}+E_{11}^{\text{UAV}}+E^{\text{UAV}}),      \label{C-e}\\
		&\text{var.}~ x,e.  \nonumber 
	\end{align}%
\end{subequations}
We can observe that the mapping $H(x,e)$ rises monotonically as variables $x$ and $e$	grow separately. When $H(x,e)$ attains its maximum value, the objective term of subproblem $\mathbf{L}\mbox{-}\mathbf{f1}$ reaches its minimal value, and the corresponding value of $x$ delivers the optimal decision variable.
Based on the constraints \eqref{Cpn2-bf'}, \eqref{C-x} and \eqref{C-e}, we define the normal cone set $\mathcal X$ and conormal cone set $\mathcal Y$	as presented below:
\begin{equation}
	\begin{aligned} \label{set-X}
		\mathcal X = \{(x,e)|\text{constraints} ~\eqref{C-x} ~\text{and}~ \eqref{C-e} \}.
	\end{aligned}
\end{equation}
\begin{equation}
	\begin{aligned} \label{set-Y}
		\mathcal Y = \{(x,e)|\text{constraint} ~\eqref{Cpn2-bf'} \}.
	\end{aligned}
\end{equation}
Exploiting the feature of problem $\mathbf{L}\mbox{-}\mathbf{f2}$ in Proposition 1, we put forward Algorithm 1 built upon polyblock approximation from references \cite{10726693}, \cite{8187134}. This algorithm aims to search for the optimal pair $(x^\ast,e^\ast)$ that minimizes the value of $H(x,e)$.

\begin{algorithm}[htbp]
	\caption{To Solve Problem $\mathbf{L}\mbox{-}\mathbf{f2}$ and Obtain Optimal $(x^\ast,e^\ast)$}
	\label{Algorithm_1}
	\begin{algorithmic}[1]		
					\STATE \textbf{Initialize} the vertex set $V_0 =\{(x^{\max},e^{\max})\}$. Set the current best value $CBV_0=-\infty$ and the current best solution $CBS_0$ as an empty set. Set the iteration number $i=0$. Set $\nu$ as a very small positive number.
					\REPEAT
					\STATE Update $i = i + 1$
					\STATE Find the vertex $z_i = \arg \max\{H(x,e)|(x,e)\in V_{i-1}\}$ from $V_{i-1}$.
					\STATE Use bisection search to find the projection of $z_i$ on the boundary of $\mathcal X$, which is denoted as $\pi_{\mathcal X}(z_i)$.
					\IF {$|\pi_{\mathcal X}(z_i) - z_i| \le \nu$ }
					\STATE  Update $CBS_i=z_i$ and $CBV_i=H(z_i)$.
					\ELSE
					\IF {$\pi_{\mathcal X}(z_i)\in \mathcal X \cap \mathcal Y$ and $H(\pi_{\mathcal X}(z_i)) \ge CBV_{i-1}$}
					\STATE Update $CBS_i = \pi_{\mathcal X}(z_i)$ and $CBV_i = H(\pi_{\mathcal X}(z_i))$.
					\ELSE
					\STATE Update $CBS_i = CBS_{i-1}$ and $CBV_i = CBV_{i-1}$.
					\ENDIF
					\STATE Set $z_{\text{tem}}=\pi_{\mathcal X}(z_i)$ and update $V_i=\{V_{i-1} \backslash z_i\}\cup \{(z_i(1),z_\text{tem}(2)),(z^n(1),z_\text{tem}(2))\}$.
					\STATE Remove from $V_i$ the improper vertices and the vertices $\{v \in V_i|v \notin \mathcal Y\}$.
					\ENDIF
					\UNTIL {$|H(z_i)-CBV_i|<\nu$}
					\STATE \textbf{Output:} the optimal solution $(x^\ast,e^\ast) = CBS_j$ for problem $\mathbf{L}\mbox{-}\mathbf{f2}$. Notice that $x^\ast$ is also the solution for problem $\mathbf{L}\mbox{-}\mathbf{f1}$.
		\end{algorithmic}
\end{algorithm} 

\subsection{Proposed Algorithm for Solving Problem $\mathbf{P2}\mbox{-}\mathbf{L}\mbox{-}\mathbf{w}$}
To solve problem $\mathbf{P2}\mbox{-}\mathbf{L}\mbox{-}\mathbf{w}$, we first define $\textbf{W} \triangleq \textbf{w}\textbf{w}^H$. Here, $\textbf{W}$ stands for a positive semi-definite matrix satisfying $\textbf{W} \succeq 0$ with unit rank, and we also introduce $\textbf{H} \triangleq \textbf{h}\textbf{h}^H$.
With this definition, constraint \eqref{Dtot} can be rewritten as
\begin{equation}
	\begin{aligned}
		\label{rwDtot}
		\textbf{Tr}(\textbf{H} \textbf{W}) \ge (e^{\frac{\sum_{n=1}^{N}D_n + D^{\text{target}}}{B^{\text{BS}} T_2}}-1) \sigma_{\text{BS}}^2.
	\end{aligned}
\end{equation}
Then, problem $\mathbf{P2}\mbox{-}\mathbf{L}\mbox{-}\mathbf{w}$ can be equivalently reformulated as follows:
\begin{subequations} \label{B-w'}%
	\begin{align}
		\textbf{$\mathbf{L}\mbox{-}\mathbf{w1}$:}& ~  \min  ~ E^\text{tot} \nonumber \\
		&\text{s.t.} ~\text{constraint}~\eqref{rwDtot},\nonumber \\
		&\quad ~~ \textbf{Tr}(\textbf{W}) \le p^{\text{UAV},\max},      \label{CW-bw'}\\
		&\quad ~~ \text{Rank}(\textbf{W}) = 1,      \label{R-CW-bw'}\\
		&\text{var.}~ \textbf{W}.  \nonumber 
	\end{align}%
\end{subequations}
Then, we employ the semidefinite relaxation (SDR) technique and temporarily ignore the Rank-1 constraint shown in formula \eqref{R-CW-bw'}. 
This treatment enables us to resolve the subproblem $\mathbf{L}\mbox{-}\mathbf{w1}$ via iterative computation over the subsequent optimization formulation.
\begin{subequations} \label{B-w''}%
	\begin{align}
		\textbf{$\mathbf{L}\mbox{-}\mathbf{w2}$:}& ~  \min  ~ E^\text{tot} \nonumber \\
		&\text{s.t.} ~\text{constraint}~\eqref{rwDtot},\nonumber \\
		&\quad ~~ \textbf{Tr}(\textbf{W}) \le p^{\text{UAV},\max},      \label{CW-bw'}\\
		&\text{var.}~ \textbf{W}.  \nonumber 
	\end{align}%
\end{subequations}
Next, we analyze the convex property of the relaxed problem $\mathbf{L}\mbox{-}\mathbf{w2}$ with respect to matrix $\textbf{W}$ and summarize the conclusion as Proposition 2. 
Full derivations to verify Proposition 2 are provided within Appendix B.

\noindent\textbf{Proposition 2.} The subproblem $\mathbf{L}\mbox{-}\mathbf{w2}$ belongs to the class of strictly convex optimization formulations with respect to matrix $\textbf{W}$.

From the conclusion of Proposition 2, we can efficiently handle the subproblem $\mathbf{L}\mbox{-}\mathbf{w2}$ via mature convex optimization solvers such as CVX. We conduct iterative computation for $\mathbf{L}\mbox{-}\mathbf{w2}$ and mark the converged output matrix as $\widetilde{\textbf{W}}$. Then, we construct a Rank-1 matrix solution based on $\widetilde{\textbf{W}}$ to acquire the optimal solution of subproblem $\mathbf{P2}\mbox{-}\mathbf{L}\mbox{-}\mathbf{w}$, which are presented below:
\begin{equation}
	\begin{aligned}
			\label{w*}
			\textbf{w}^\ast= (\textbf{h}^H \widetilde{\textbf{W}} \textbf{h})^{-\frac{1}{2}} \widetilde{\textbf{W}} \textbf{h}.
		\end{aligned}
\end{equation}
The details for solving problem $\mathbf{P2}\mbox{-}\mathbf{L}\mbox{-}\mathbf{w}$ are presented in Algorithm 2.
\begin{algorithm}[htbp]
	\caption{To Solve Problem $\mathbf{P2}\mbox{-}\mathbf{L}\mbox{-}\mathbf{w}$ and obtain $\textbf{w}^\ast$}
	\label{Algorithm_2}
	\begin{algorithmic}[1]		
					\STATE \textbf{Initialize} $\textbf{W}^{0}$. Set the iteration number $t$=0. Set $\Delta$ as a very small positive number.
					\REPEAT
					\STATE Update t = t+1.
					\STATE Solve problem $\mathbf{L}\mbox{-}\mathbf{w2}$ with $\textbf{W}^{t-1}$ by CVX and obtain $\textbf{W}^{t}$
					\UNTIL{Convergence.}
					\STATE Calculate $\textbf{w}^\ast$ according to Eq.\eqref{w*}.
					\STATE \textbf{Output:} the solution $\textbf{w}^\ast$ for problem $\mathbf{P2}\mbox{-}\mathbf{L}\mbox{-}\mathbf{w}$.
		\end{algorithmic} 
\end{algorithm} 

\subsection{Proposed Algorithm for Solving Problem $\mathbf{P2}\mbox{-}\mathbf{L}\mbox{-}\mathbf{s}$}
To solve problem $\mathbf{P2}\mbox{-}\mathbf{L}\mbox{-}\mathbf{s}$, we define $\textbf{X} \triangleq \textbf{A}_s^H \textbf{u}$, $\textbf{Y} \triangleq \textbf{A}_c^H \textbf{u}$ and $Z \triangleq \sigma_0^2 \textbf{u}^H \textbf{u}$. Then, constraints \eqref{CTtot}, \eqref{CQ} and \eqref{CT1} can be rewritten as
\begin{equation}
	\begin{aligned}
		\label{rew-CTtot}
		\textbf{X}^H \textbf{S}_0 \textbf{X} \le C_1 (\textbf{Y}^H \textbf{S}_0 \textbf{Y} + Z),
	\end{aligned}
\end{equation}
\begin{equation}
	\begin{aligned}
		\label{rew-CQ}
		\textbf{X}^H \textbf{S}_0 \textbf{X} \ge C_2 (\textbf{Y}^H \textbf{S}_0 \textbf{Y} + Z),
	\end{aligned}
\end{equation}
\begin{equation}
	\begin{aligned}
		\label{rew-CT11}
		\textbf{X}^H \textbf{S}_0 \textbf{X} \le C_3 (\textbf{Y}^H \textbf{S}_0 \textbf{Y} + Z),
	\end{aligned}
\end{equation}
where 
\begin{equation}
	\begin{aligned}
		\label{C1}
		C_1 = \frac{1}{2 \tau B^{\text{UAV}}} (2^{(T^{\text{tot},\max}-T_{11}-T_2) (\frac{2 f^{\text{UAV}} \tau }{T_{11} C^{\text{UAV}} \delta} ) }-1)),
	\end{aligned}
\end{equation}
\begin{equation}
	\begin{aligned}
		\label{C2}
		C_2 = \frac{1}{2 \tau B^{\text{UAV}}}  (2^{\frac{2 \tau Q^{\text{req}}}{\delta}}-1),
	\end{aligned}
\end{equation}
\begin{equation}
	\begin{aligned}
		\label{C3}
		C_3 = \frac{1}{2 \tau B^{\text{UAV}}}  (2^{\frac{2 (T_{11}^{\max}-T_{11}) f^{\text{UAV}} \tau }{T_{11} C^{\text{UAV}} \delta}}-1),
	\end{aligned}
\end{equation}
The objective function together with constraints \eqref{Dtot} and \eqref{Cpn2} contains non-convex terms, so
we apply SCA to solve them and obtain a more tractable form.
We define $\theta(\textbf{S}_0) = (\mathbf{g}_n^{t2})^H \Omega^{-1} \mathbf{g}_n^{t2}$, 
$\gamma(\textbf{S}_0) = 2 \tau B^{\text{UAV}} \frac{\textbf{X}^H \textbf{S}_0 \textbf{X}}{\textbf{Y}^H \textbf{S}_0 \textbf{Y} + Z}$, 
$\psi(\textbf{S}_0) = \log(1+\gamma(\textbf{S}_0))$, 
$\eta_1(\textbf{S}_0)=\frac{\sum_{n=1}^{N} D_{n}}{T_{11}+ \frac{T_{11} C^{\text{UAV}} \delta}{f^{\text{UAV}} 2 \tau}  \psi(\textbf{S}_0)}$, and
$\eta_2(\textbf{S}_0)=\frac{\sum_{\Theta(m) > \Theta(n)} D_m}{T_{11}+ \frac{T_{11} C^{\text{UAV}} \delta}{f^{\text{UAV}} 2 \tau}  \psi(\textbf{S}_0)}$.
With these auxiliary mappings defined, we can restate the objective function and constraints \eqref{Dtot}, \eqref{Cpn2} in the following equivalent form:
\begin{equation}
	\begin{aligned}
		\label{rew-Dtot}
		f_4(\textbf{S}_0)=\sum_{n=1}^{N}D_n +\alpha T_{11} \psi(\textbf{S}_0) \le R^{\text{UAV}} T_2, 
	\end{aligned}
\end{equation}
\begin{equation}
	\begin{aligned}
		\label{rew-Cpn2}
		f_1(\textbf{S}_0)=\frac{1}{\theta(\textbf{S}_0)} (2^{\eta_1(\textbf{S}_0)} - 2^{\eta_2(\textbf{S}_0)}) \le p_n^{\text{USV},\max}, 
	\end{aligned}
\end{equation}
\begin{equation}
	\small
	\begin{aligned}
		\label{rew-objfun1}
		E^{\text{tot}} = & \sum_{n=1}^{N} f_1(\textbf{S}_0) (T_{11}+\frac{T_{11} C^{\text{UAV}} \delta}{f^\text{UAV} 2 \tau} \log(1+\gamma(\textbf{S}_0))) +\text{Tr}(\textbf{S}_0) T_{11} \nonumber \\
		&+\omega^\text{UAV} T_{11} C^{\text{UAV}} (f^\text{UAV})^2 \frac{\delta}{2 \tau} \log(1+\gamma(\textbf{S}_0)) +\text{Tr}(\textbf{w} \textbf{w}^H) T_2  \nonumber \\
	\end{aligned}
\end{equation}
\begin{equation}
	\small
	\begin{aligned}
		\label{rew-objfun2}
		&+\frac{1}{2}v w^{\text{UAV}} (T_{11} +T_2 +\frac{T_{11} C^{\text{UAV}} \delta}{f^\text{UAV} 2 \tau} \log(1+\gamma(\textbf{S}_0)) ).  \\
	\end{aligned}
\end{equation}
By analyzing the convex approximation property of the transformed non-convex expressions, we draw the conclusion presented in Proposition 3. 
Full derivations for verifying Proposition 3 are placed in Appendix C.

\noindent\textbf{Proposition 3.} We can substitute the original objective function together with constraints \eqref{Dtot} and \eqref{Cpn2} by the convex objective term and convex constraints listed below
\begin{equation}
	\begin{aligned}
		\label{con-rew-Dtot}
		f_4(\textbf{S}^{(k)}) + \bigtriangledown f_4(\textbf{S}^{(k)}) (\textbf{S}_0-\textbf{S}^{(k)}) \le R^{\text{UAV}} T_2, 
	\end{aligned}
\end{equation}
\begin{equation}
	\begin{aligned}
		\label{con-rew-Cpn2}
		f_1(\textbf{S}^{(k)}) + \bigtriangledown f_1(\textbf{S}^{(k)}) (\textbf{S}_0-\textbf{S}^{(k)}) \le p_n^{\text{USV},\max}, 
	\end{aligned}
\end{equation}
\begin{equation}
	\begin{aligned}
		\label{con-rew-objfun}
		E^{\text{tot}} \approx E^{\text{tot}}(\textbf{S}^{(k)}) + \bigtriangledown E^{\text{tot}}(\textbf{S}^{(k)}) (\textbf{S}_0-\textbf{S}^{(k)}),
	\end{aligned}
\end{equation}

Therefore, problem $\mathbf{P2}\mbox{-}\mathbf{L}\mbox{-}\mathbf{s}$ can be restated into a standard convex optimization formulation presented below, and we can also use CVX to solve this problem.
\begin{subequations} \label{B-s'}%
	\begin{align}
		\textbf{$\mathbf{L}\mbox{-}\mathbf{s1}$:}& ~  \min  ~ E^\text{tot} \nonumber \\
		&\text{s.t.} ~\text{constraints}~\eqref{CS0},~\eqref{rew-CTtot},~\eqref{rew-CQ},~\eqref{rew-CT11},~\eqref{con-rew-Cpn2}, \nonumber \\ 
		&\text{var.}~ \textbf{S}_0.  \nonumber 
	\end{align}%
\end{subequations}

\subsection{Proposed Algorithm for Solving Problem $\mathbf{P2}\mbox{-}\mathbf{U}$}
In problem $\mathbf{P2}\mbox{-}\mathbf{U}$, by performing the first-order Taylor expansion of $T_{11}$ in $f(T_{11}) = 2^{\frac{\sum_{n=1}^{N} D_n}{T_{11}(1+\frac{Q C^{\text{UAV}}}{f^\text{UAV}})B^{\text{UAV}}}}- 2^{\frac{ \sum_{\Theta(m) > \Theta(n)} D_m}{T_{11}(1+\frac{Q C^{\text{UAV}}}{f^\text{UAV}})B^{\text{UAV}}}}$ using the SCA method, we obtain $f(T_{11}) \approx f(T^{(k)}_{11})+\bigtriangledown f(T^{(k)}_{11})(T_{11}-T^{(k)}_{11}) \approx f(T^{(k)}_{11})+(-\frac{\ln2 \sum_{n=1}^{N} D_n }{(1+\frac{Q C^{\text{UAV}}}{f^\text{UAV}})B^{\text{UAV}}(T^{(k)}_{11})^2} 2^{\frac{ \sum_{n=1}^{N} D_n }{(1+\frac{Q C^{\text{UAV}}}{f^\text{UAV}})B^{\text{UAV}}T^{(k)}_{11}}} + \frac{\ln2 \sum_{\Theta(m) > \Theta(n)} D_m }{(1+\frac{Q C^{\text{UAV}}}{f^\text{UAV}})B^{\text{UAV}}(T^{(k)}_{11})^2} 2^{\frac{ \sum_{\Theta(m) > \Theta(n)} D_m }{(1+\frac{Q C^{\text{UAV}}}{f^\text{UAV}})B^{\text{UAV}}T^{(k)}_{11}}}) (T_{11}-T^{(k)}_{11})$. 
Problem $\mathbf{P2}\mbox{-}\mathbf{U}$ can be transformed into convex problem $\mathbf{U}\mbox{-}\mathbf{T1}$, which can be efficiently solved using the CVX optimization toolbox.
\begin{subequations} \label{p2-T'}%
	\small
	\begin{align}
		\textbf{$\mathbf{U}\mbox{-}\mathbf{T1}$:} ~ & \min  ~ \sum_{n=1}^{N} \big((\mathbf{g}_n^{t2})^H \Omega^{-1} \mathbf{g}_n^{t2}\big)^{-1}  f(T_{11}) (T_{11}+\frac{Q T_{11} C^{\text{UAV}}}{f^\text{UAV}}) \nonumber \\
		&+\text{Tr}(\textbf{S}_0) T_{11} +\omega^\text{UAV} Q T_{11} C^{\text{UAV}} (f^\text{UAV})^2 +\text{Tr}(\textbf{w} \textbf{w}^H) T_2  \nonumber \\
		&+\frac{1}{2}v w^{\text{UAV}} (T_{11}+\frac{Q T_{11} C^{\text{UAV}}}{f^\text{UAV}} +T_2) \nonumber \\
		&\text{s.t.} ~\text{constraints}~\eqref{Dtot},~\eqref{CTtot},~\eqref{CT1},~\eqref{CT2},\nonumber \\ 
		&\quad ~~ \big((\mathbf{g}_n^{t2})^H \Omega^{-1} \mathbf{g}_n^{t2}\big)^{-1} f(T_{11}) \le p_n^{\text{USV},\max},      \label{T'-Cpn2} \tag{57a}\\
		&\text{vars.}~ T_{11},~T_2.  \nonumber 
	\end{align}%
\end{subequations}

Finally, to solve the original problem $\mathbf{P1}$, we execute Algorithm 3 and problem $\mathbf{P2}\mbox{-}\mathbf{U}$ alternately until their respective solutions keep unchanged. The details for solving problem $\mathbf{P2}\mbox{-}\mathbf{L}$ are presented in Algorithm 3.
\begin{algorithm}[htbp]
	\caption{BCD-based Algorithm to Find the optimal Solution of Problem $\mathbf{P2}\mbox{-}\mathbf{L}$}
	\label{Algorithm_3}
	\begin{algorithmic}[1]		
			\REPEAT
			\STATE Using Algorithm 1 to slove problem $\mathbf{P2}\mbox{-}\mathbf{L}\mbox{-}\mathbf{f}$ and obtain the value of \( f^{\text{UAV}} \).
			\STATE Using Algorithm 2 to slove problem $\mathbf{P2}\mbox{-}\mathbf{L}\mbox{-}\mathbf{w}$ and obtain the value of \( \boldsymbol{w} \).
			\UNTIL the solutions of problems $\mathbf{P2}\mbox{-}\mathbf{L}\mbox{-}\mathbf{f}$, $\mathbf{P2}\mbox{-}\mathbf{L}\mbox{-}\mathbf{w}$ and $\mathbf{P2}\mbox{-}\mathbf{L}\mbox{-}\mathbf{s}$ keep unchanged.
		\end{algorithmic} 
	\end{algorithm} 

We next analyze the computational complexity of the proposed algorithm. Algorithm 1 is invoked in Step 2 of Algorithm 3 to solve problem $\mathbf{P2}\mbox{-}\mathbf{L}\mbox{-}\mathbf{f}$ and its complexity is dominated by vertex set update steps (Lines 4, 14, and 15) and bisection search (Line 5). Each vertex set update step has a complexity of $O(|\mathcal{V}|)$ where $|\mathcal{V}|$ denotes the size of the vertex set in Algorithm 1. The bisection search has a complexity of $O\left(\log \frac{1}{\epsilon}\right)$ with $\epsilon$ as the search precision \cite{10955684}. Algorithm 2 is invoked in Step 3 of Algorithm 3 to solve problem $\mathbf{P2}\mbox{-}\mathbf{L}\mbox{-}\mathbf{w}$ and its complexity mainly arises from interior-point method-based solution for the beamforming subproblem, yielding a complexity of $O(N_t^3)$ where $N_t$ is the number of UAV transmit antennas. The SCA step in Step 4 of Algorithm 3 targets problem $\mathbf{P2}\mbox{-}\mathbf{L}\mbox{-}\mathbf{s}$ and has a complexity of $O(N_t^3)$ per iteration. Problem $\mathbf{P2}\mbox{-}\mathbf{U}$ focuses on phase duration optimization and involves two scalar variables which are durations of the two phases, so its complexity is $O(1)$ as a constant term per iteration. Let $K$ denote the number of iterations for Algorithm 3 to converge. The overall complexity of Algorithm 3 is thus $O\left(K \cdot \left(|\mathcal{V}| + \log \frac{1}{\epsilon} + N_t^3\right)\right)$ where the per-iteration cost is dominated by the $O(N_t^3)$ term from beamforming and sensing signal optimization. The stopping criterion for Algorithm 3 is defined to terminate when the relative change in the total energy consumption $E^{\text{tot}}$ between consecutive iterations is less than $10^{-3}$. The constant complexity of problem $\mathbf{P2}\mbox{-}\mathbf{U}$ is combined into the asymptotic analysis. Accordingly, the overall computational complexity of the proposed algorithm is $O\left(K \cdot \left(|\mathcal{V}| + \log \frac{1}{\epsilon} + N_t^3\right)\right)$.

\section{Simulation Results} \label{Numerical Results}
This section presents numerical simulations to evaluate the performance of our proposed algorithm and overall system framework.
Referring to the IoT network parameters in the literature [35], 
we set the UAV platform to equip $N_t = 10$ transmitting antennas and $N_r = 15$ receiving antennas, the target sensing node is placed at the azimuth angle $\theta_0 = 0^{\circ}$, while all clutter scatterers correspond to angular positions $\{-50^{\circ}, -30^{\circ}, 30^{\circ}, 50^{\circ}\}$. 
The horizontal separation between the BS and UAV is fixed at $500$ m, and all USVs are randomly distributed within a $200$ m radius centered on the UAV.
We model the wireless links between USVs and the UAV, as well as the connections linking the UAV and SBS, as Rayleigh fading channels with standard path loss attenuation, and we configure core system parameters as $B^{\text{UAV}} = B^{\text{BS}} = 10$ MHz for communication bandwidth, $\delta = 0.01$ for the duty cycle factor, $\tau = 2 \times 10^{-6}$ s for pulse time length, $C^{\text{UAV}} = 1 \times 10^3$ cycles/bit for the UAV’s per-bit CPU computation overhead, $\omega^{\text{UAV}} = 1 \times 10^{-28}$ for the UAV GPU power efficiency coefficient, $\sigma_{\text{BS}}^2 = \sigma_{0}^2$ as -$174$ dBm/Hz for background noise power, $1$ W as the peak transmit power limit, data payload sizes ranging from $1$ Mbits to $2$ Mbits, a maximum Phase-I time budget $T_1^{\max}$ s and a maximum Phase-II time budget $T_2^{\max}$ s. The self-interference channel matrix at the UAV follows the expression \([\mathbf{H}_{\text{SI}}]_{m,n} = \sqrt{\aleph_{m,n}} e^{-j2\pi \frac{d_{m,n}}{\lambda}}\), where \(\aleph_{m,n} = -110\) dB stands for the self-interference channel gain, and \(\hat{d}_{m,n}\) denotes the spatial gap separating the \(m\)-th receive antenna and the \(n\)-th transmit antenna. 

\subsection{Convergence and Optimality Performance}
This subsection carries out verification on the convergence performance and optimality of our proposed optimization algorithm.
We consider four scenarios, and the number of USVs is configured as $3$, $5$, $7$ and $9$. 

\begin{figure}[h]
	\centering
	\includegraphics[scale=0.5]{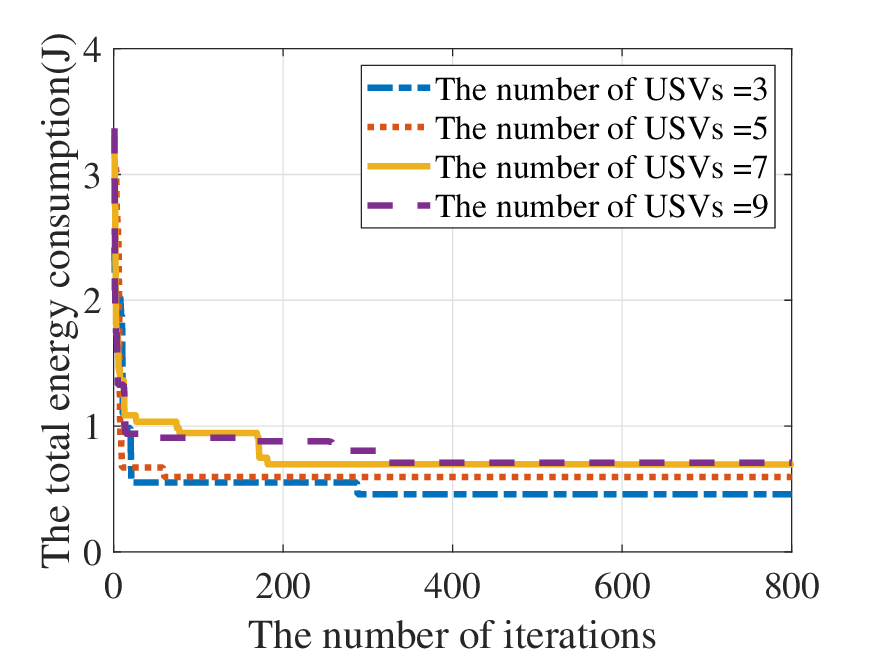} 
	\caption{The convergence performance for different numbers of USVs.}
	\label{shoulian} 
\end{figure}

Fig. \ref{shoulian} illustrates the convergence performance of our proposed optimization algorithm under varying quantities of USVs. We can observe that the algorithm converges within $200$ iterations when the USV count equals $5$ or $7$, while approximately $250$ iterations are required to reach convergence for cases with $3$ or $9$ USVs
Moreover, total system energy expenditure rises as the number of USVs increases.
This is because, under the same latency constraint, a larger number of USVs results in a higher volume of data to be transmitted, thereby requiring more energy to satisfy the offloading demand.

\begin{figure}[h]
	\centering
	\includegraphics[scale=0.25]{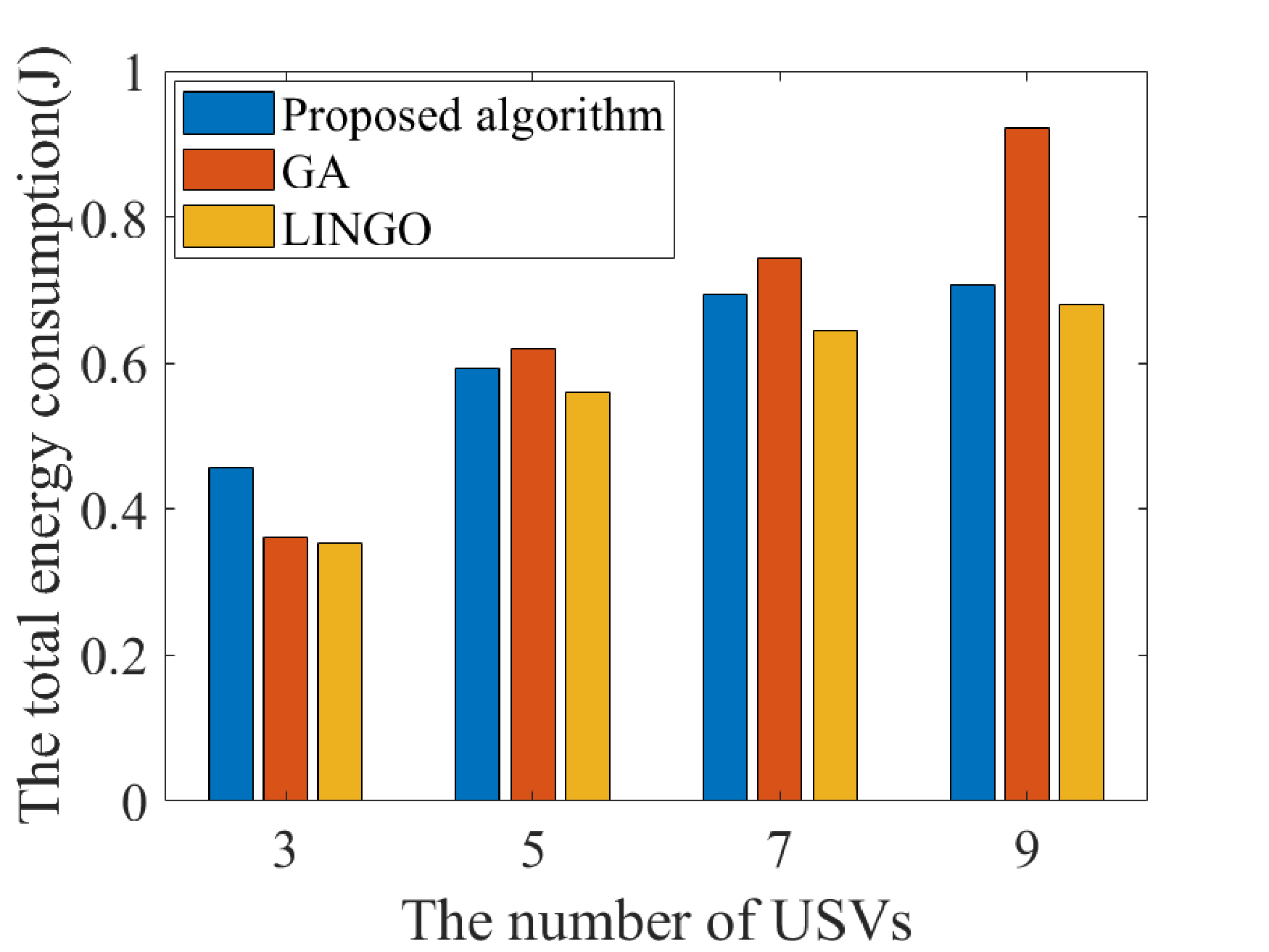} 
	\caption{The accuracy of the proposed algorithms in comparison with GA and LINGO.}
	\label{ourandGAE} 
\end{figure}

\begin{figure}[h]
	\centering
	\includegraphics[scale=0.24]{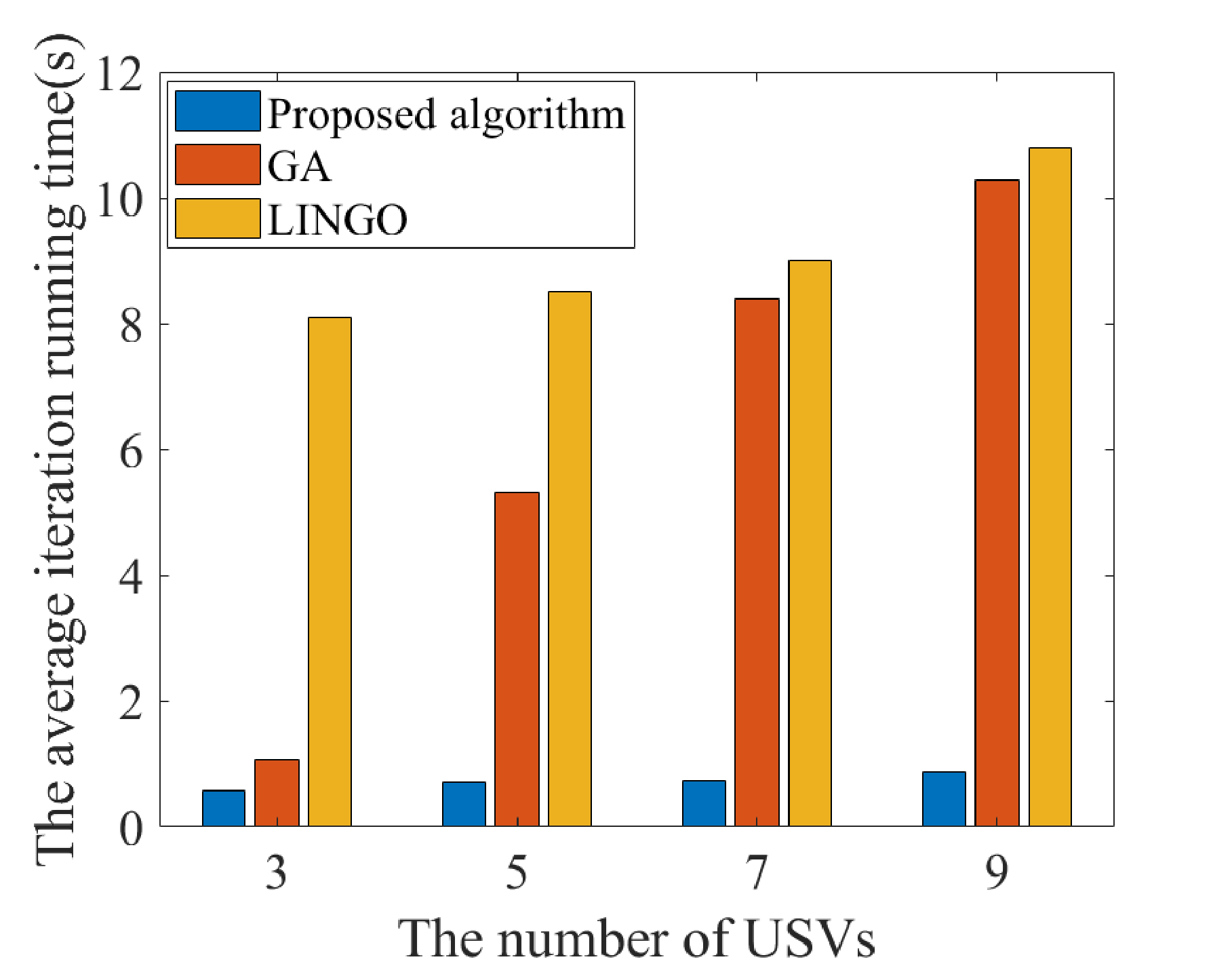} 
	\caption{The average running time per time slot of the proposed, GA algorithms and LINGO.}
	\label{ourandGAT} 
\end{figure}

In Fig. \ref{ourandGAE}, we compare the total system energy consumption of the proposed algorithm, the genetic algorithm (GA)\footnote{GA is the comparison benchmark since it is a well-established heuristic for complex optimization, capable of exploring solution spaces to yield expected feasible solutions, which enables valid performance verification of the proposed algorithm.}, and linear interactive and general optimizer (LINGO)\footnote{LINGO is a professional optimization tool that can provide high-precision solutions for nonlinear programming problems, so we adopt it as a benchmark to verify the optimality of the proposed algorithm.} under different numbers of USVs. The proposed algorithm outperforms the GA, with an average deviation in total energy consumption of no more than $8$\%, and stays within $8.72$\% of the result obtained by LINGO. In Fig. \ref{ourandGAT}, 
we carry out comparisons of average runtime among our proposed algorithm, GA and LINGO. Numerical outcomes reveal that the average runtime of our proposed algorithm is over $7$ times shorter than that of GA and more than $11$ times lower compared with LINGO.

\subsection{Performance of the Proposed Scheme}
This subsection assesses the overall performance of our proposed scheme via cross-checks against multiple benchmark schemes.

\begin{figure}[h]
	\centering
	\includegraphics[scale=0.26]{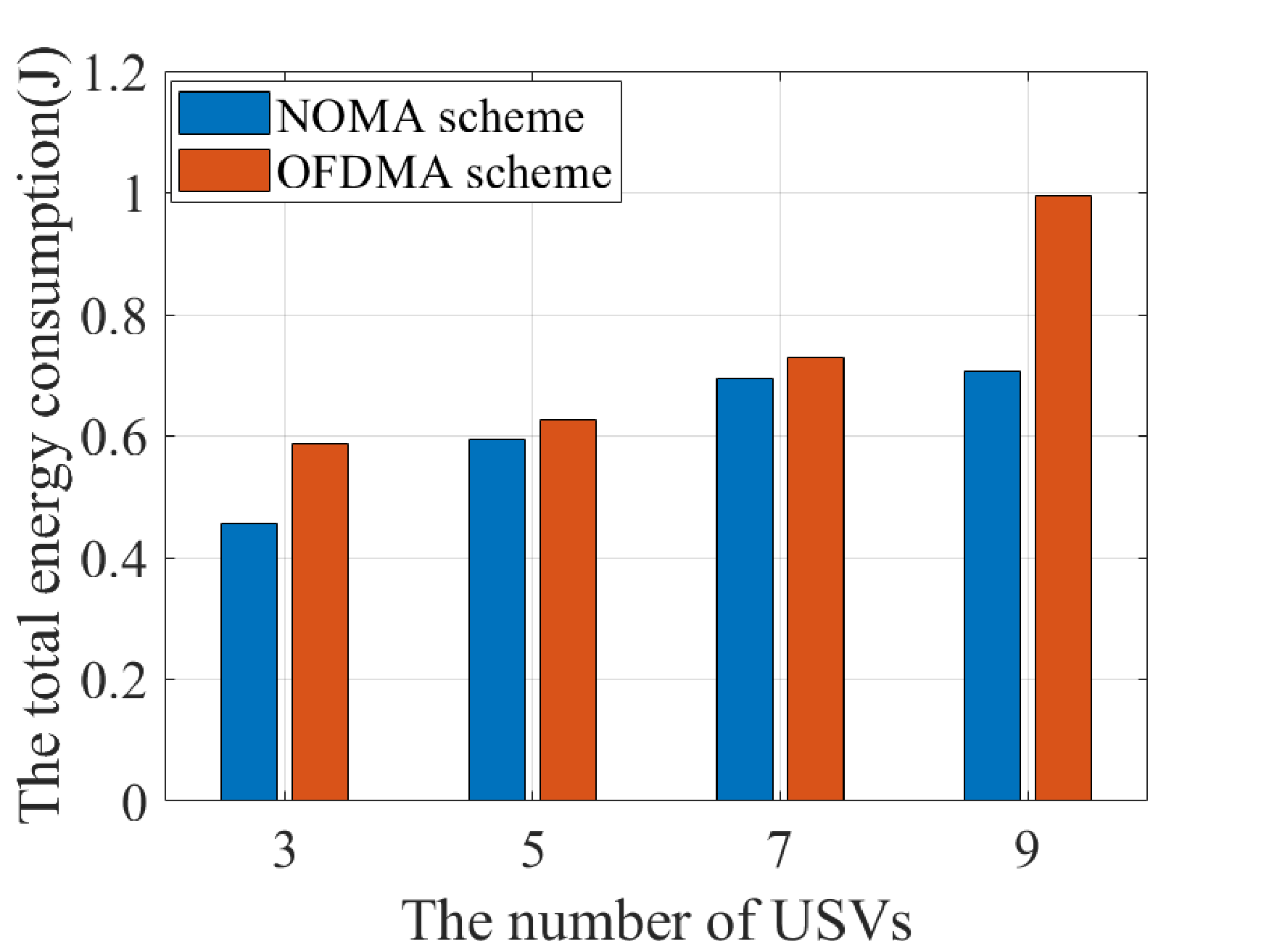} 
	\caption{The total energy consumption obtained by NOMA, and OFDMA at the different USV densities.}
	\label{NOMAandFDMAE} 
\end{figure}

Fig. \ref{NOMAandFDMAE} delivers a comparative analysis of overall system energy consumption between our proposed NOMA-based transmission scheme and the Orthogonal Frequency Division Multiple Access (OFDMA) scheme.
The results show that our NOMA-based scheme can reduce the system energy consumption by $19.71$\% compared to the OFDMA scheme. Since NOMA superimposes multiple USV signals within the same time-frequency resource block and employs SIC technology, it effectively improves spectral efficiency and thus reduces system energy consumption. It implies that battery-limited USV fleets can enjoy prolonged endurance and an extended operating range.

\begin{figure}[h]
	\centering
	\includegraphics[scale=0.39]{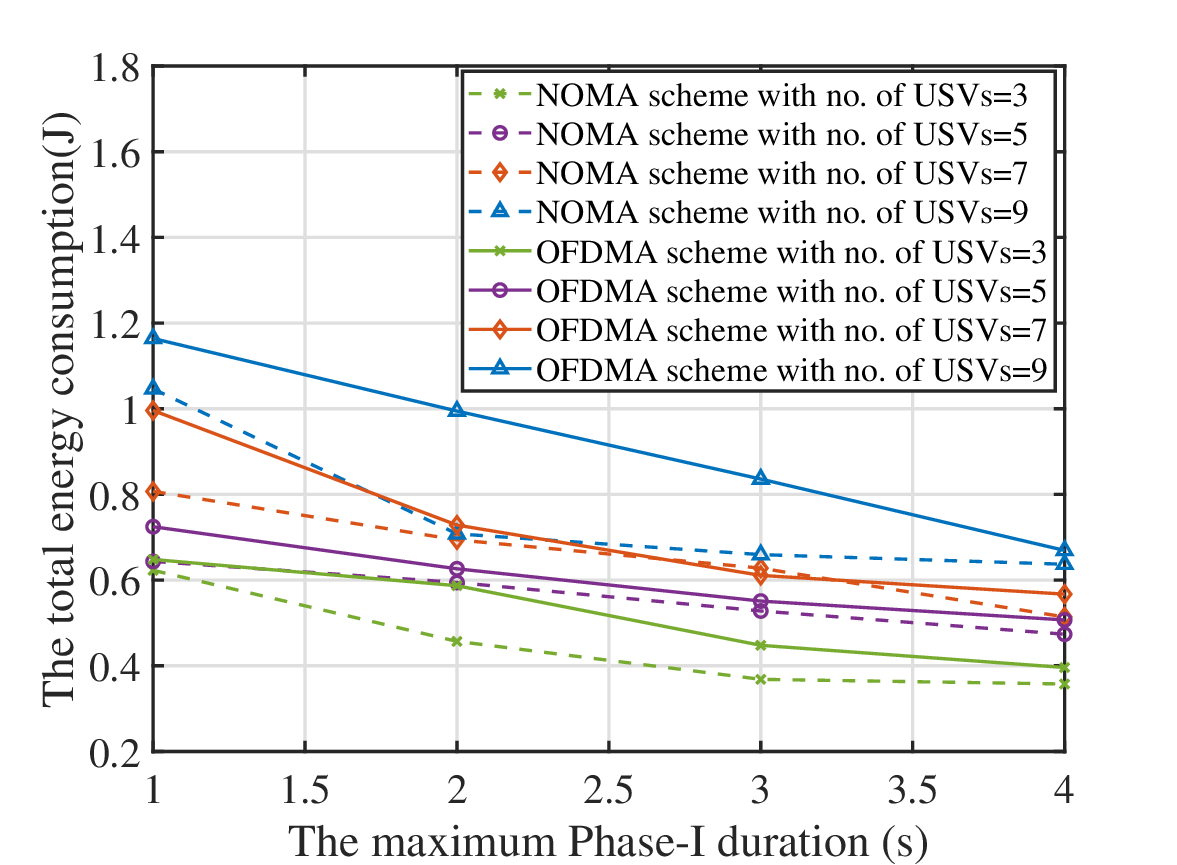}
	\caption{The total energy consumption under different maximum Phase-I duration.}
	\label{NOMAandFDMAT1max} 
\end{figure}

Fig. \ref{NOMAandFDMAT1max} further illustrates that the overall energy consumption decreases as the maximum allowable duration in Phase I increases. Moreover, as the number of USVs increases from $3$ to $9$, the proposed NOMA-based scheme consistently achieves lower energy consumption than that of the OFDMA-based scheme. Specifically, the NOMA can reduce the total energy consumption by $13.5$\% on average, in comparison with the OFDMA. This is because that a longer allowable duration in Phase I lets each USV lower its transmit power, so the total energy consumption decreases.

\begin{figure}[h]
	\centering
	\includegraphics[scale=0.52]{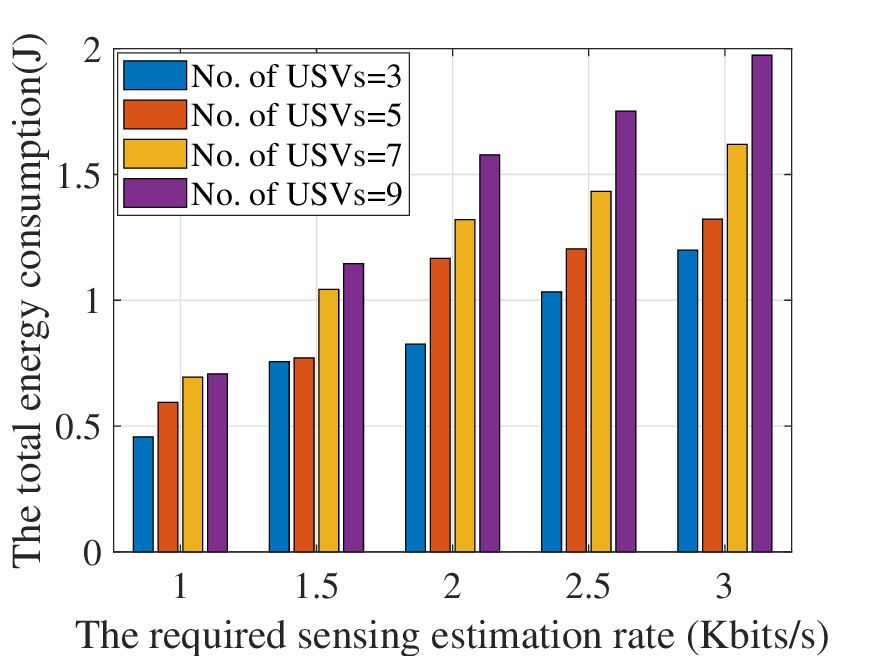} 
	\caption{The total energy consumption under different required sensing estimation rates.}
	\label{reqsensingestrate} 
\end{figure}

Fig. \ref{reqsensingestrate} plots the total system energy consumption with different required sensing estimation rates, when varying the number of USVs varies from $3$ to $9$. It can be seen that given the number of USVs, the total system energy consumption always increases with the increase of required sensing estimation rates.
This is because that higher sensing rate requirements lead to greater communication demands for each USV, which results in higher transmit power or longer transmission durations. It implies that the energy-aware mission planning should jointly tune sensing quality and USV density to keep the battery budget balanced. 

\begin{figure}[h]
	\centering
	\includegraphics[scale=0.52]{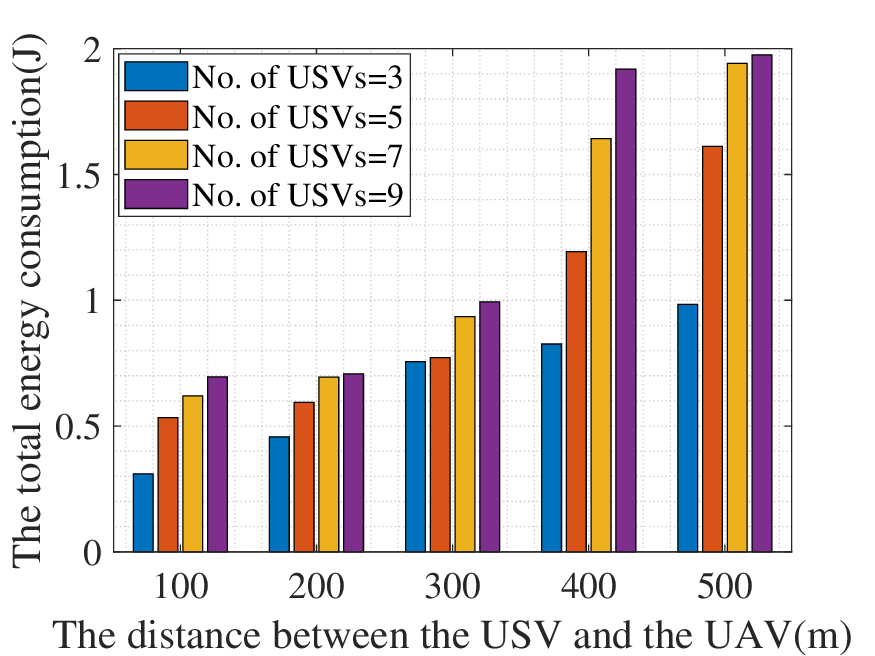} 
	\caption{The total energy consumption under different distances between the USV and UAV.}
	\label{distance} 
\end{figure}

Fig. \ref{distance} plots the total system energy consumption under different distances between the USVs and the UAV, when varying the number of USVs from $3$ to $9$. It can be seen that the total energy consumption increases with as the distance between USV and UAV increases. This is because that farther distance leads to higher path loss and degraded channel quality. It implies that the same data rate requirement should be satisfied at the cost of each USV to transmit with higher power or longer duration for the USV.

\begin{figure}[h]
	\centering
	\includegraphics[scale=0.2]{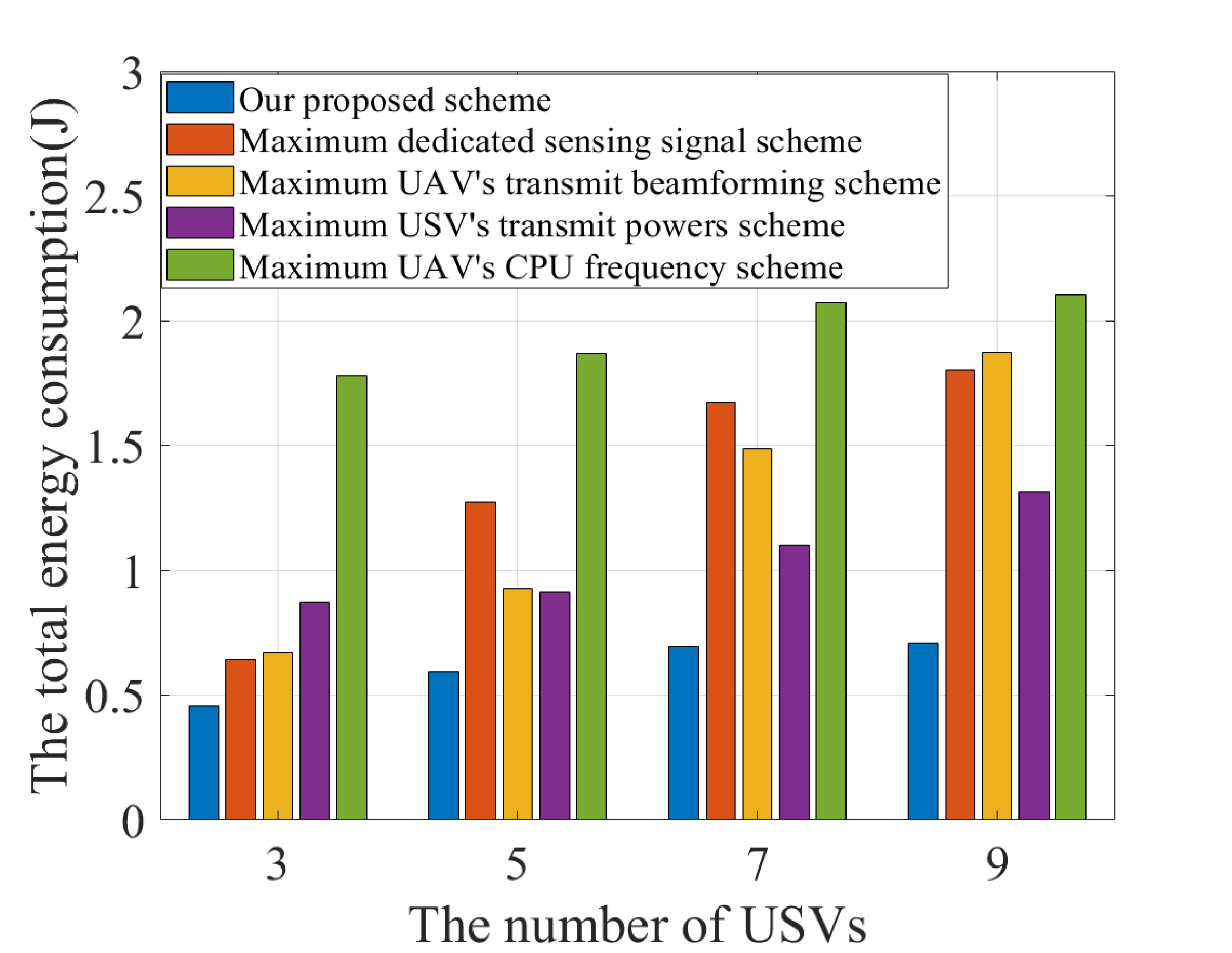}
	\caption{The total energy consumption under different optimization schemes.}
	\label{duibi} 
\end{figure}

In Fig. \ref{duibi}, we compare the proposed scheme with four benchmark schemes, the maximum dedicated sensing signal scheme, the maximum UAV's transmit beamforming scheme, the maximum USV's transmit power scheme, and the maximum UAV's CPU frequency scheme, in terms of energy consumption.  
We can find that, compared to the four baseline schemes, the proposed scheme achieves average reductions in energy consumption by factors of $1.2$, $1$, $0.7$, and $2$, respectively. 
These results demonstrate that we can significantly improve the energy efficiency and reduce the overall system energy expenditure by jointly optimizing the dedicated sensing signal, UAV's transmit beamforming, UAV's transmit powers, and UAV's CPU frequency. It implies that our proposed approach can be applied to scenarios with energy-constrained UAV fleets, and UAVs can extend their mission duration or sense more targets without extra battery power supplies.

\begin{figure}[h]
	\centering
	\includegraphics[scale=0.26]{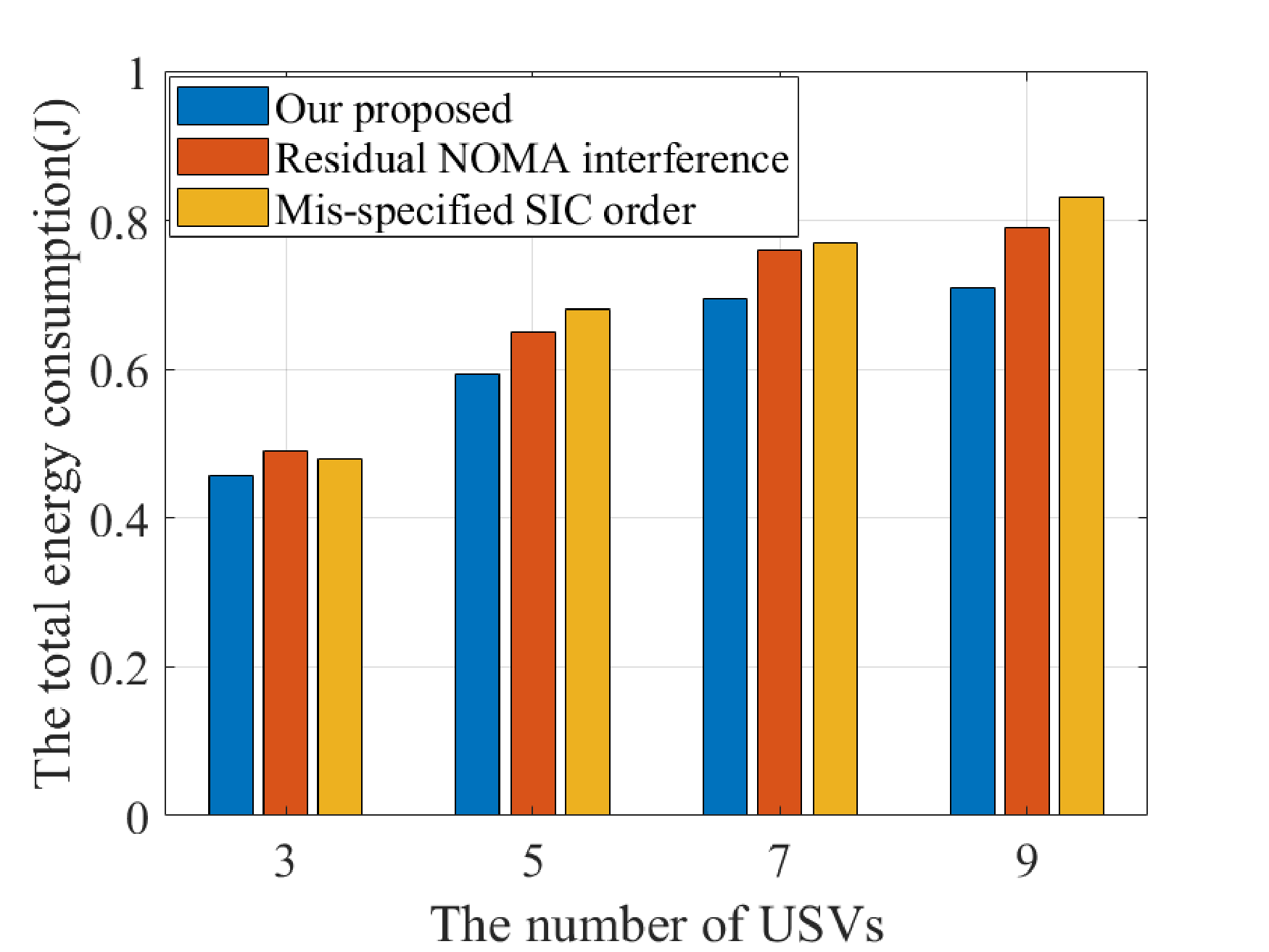} 
	\caption{The total energy consumption of our proposed scheme compared with residual NOMA interference and mis-specified SIC order} 
	\label{ourordermisandresidualNOMA} 
\end{figure}

As we know, we can simplify the solution of energy consumption minimization by assuming the perfect interference cancellation and SIC decoding order at the cost of performance degradation. Thus, in Fig. \ref{ourordermisandresidualNOMA}, we carry out numerical simulations to test the robustness of our proposed scheme under scenarios with residual NOMA interference and mismatched SIC decoding orders. Fig. \ref{ourordermisandresidualNOMA} records the overall energy consumption achieved by our proposed scheme within the two imperfect channel setups.
Numerical observations reveal that when our proposed scheme is applied to these two imperfect scenarios, it exhibits slight performance loss. Specifically, the loss is only $9.66$\% and $12.52$\% for the residual interference and the mis-specified SIC order, respectively. Since more optimal energy consumption is needed to mitigate the co-channel interference in the imperfect condition than that in the perfect condition, the performance loss is less when compared to the optimal energy consumption needed in these imperfect scenarios. This result confirms that our proposed scheme can be effectively applied to imperfect scenarios with low computational complexity.

\section{Conclusion} \label{Conclusion}
This paper investigates the energy consumption minimization problem for an ISAC system in MIoT. In this system, the UAV performs target sensing and data collection from USVs during the first phase, and subsequently forwards both the sensing and the data collected from the USVs to the SBS in the second phase. 
We formulate a joint optimization problem covering the UAV’s transmit beamforming, dedicated sensing signal design, USVs’ transmit powers, the computation power of the UAV, and the durations in the sensing and communication phases, with the objective of minimizing overall system energy consumption. 
Although the proposed joint optimization problem is non-convex, we develop a decomposition-based framework with corresponding algorithms to solve it effectively.  
Numerical outcomes verify the accuracy and efficiency of our proposed optimization algorithm.
For future research directions, we will address the limitations of this single-target sensing model and static channel assumption. We will focus on multi-target sensing scenarios to extend the applicability of the proposed ISAC scheme to complex marine environments.

\begin{appendices}
	\setcounter{equation}{0}
	\section{Proof of Proposition 1}\label{appendicesA}
	The objective function of subproblem $\mathbf{L}\mbox{-}\mathbf{f1}$ 
	falls into five components, namely 
	all USVs’ energy consumption $\sum_{n=1}^{N}E_n^{\text{USV}}$ for transmitting data, the UAV's energy consumption for sensing data $E_{11}^{\text{UAV}}$, computing data $E_{12}^{\text{UAV}}$ and data forwarding $E^{\text{UAV}}$, the UAV's energy consumption for hovering $E^{\text{UAV,hov}}$. 
	According to Eq. (12) and Eq.(29), it can be observed that $E^{\text{UAV,hov}}$ 
	exhibits an upward trend as the variable $x$ rises. 
	$E_{11}^{\text{UAV}}$ and $E^{\text{UAV}}$ are constants with respect to $x$, and $E_{12}^{\text{UAV}}$ is decreasing with respect to $x$. 
	We then prove that the cumulative USV transmission energy $\sum_{n=1}^{N}E_n^{\text{USV}}$ decreases as $x$ grows.
	
	The first-order derivative of $E_n^{\text{USV}}$ with respect to $x$ can be expressed as \renewcommand\theequation{A.\arabic{equation}}
	\begin{equation}
		\begin{aligned} \label{EnUSV'}
			\frac{\mathrm{d} E_n^{\text{USV}}}{\mathrm{d} x}
			&=C^{-1} \big[2^{\frac{A+B}{x}}(1-\ln2 \frac{A+B}{x})- 2^{\frac{B}{x}}(1-\ln2 \frac{B}{x}) \big] \\
			&=C^{-1}( h(A+B)-h(B))
		\end{aligned}
	\end{equation}
	where $A=\frac{D_n}{B^{\text{UAV}}}$, $B=\frac{\sum_{\Theta(m) > \Theta(n)} D_m}{B^{\text{UAV}}}$, $C=(\mathbf{g}_n^{t2})^H \Omega^{-1} \mathbf{g}_n^{t2}$ and $h(a)=2^{\frac{a}{x}}(1-\ln2 \frac{a}{x}), x > 0$. We can obtain the first-order derivative of $h(a)$ with respect to $a$ as 
	\begin{equation}
		\begin{aligned} \label{h(a)'}
			\frac{\mathrm{d} h(a)}{\mathrm{d} a}
			&=- a \frac{(\ln2)^2}{x^2}2^{\frac{a}{x}} < 0.
		\end{aligned}
	\end{equation}
	By (A.2), we conclude that $h(a)$ monotonically declines as the variable $a$ rises.
	Accordingly, we obtain $h(A+B)-h(B) < 0$ in (A.1). With (A.2) and $C > 0$, we can derive that $\frac{\mathrm{d} E_n^{\text{USV}}}{\mathrm{d} x} < 0$, which proves that $\sum_{n=1}^{N}E_n^{\text{USV}}$ is decreasing with respect to $x$.
	Thus, we can reorganize the objective expression of formulation $\mathbf{L}\mbox{-}\mathbf{f1}$ into
	\begin{equation}
		\begin{aligned} \label{objetotA-A}
			E^\text{tot}=E^{\text{UAV,hov}}+E_{11}^{\text{UAV}}+E^{\text{UAV}}-(E_{12}^{\text{UAV}}+\sum_{n=1}^{N}E_n^{\text{USV}}),
		\end{aligned}
	\end{equation}
	which corresponds to the gap between two functions that both rise monotonically with the increase of $x$. Therefore, Proposition 1 is proved.  $\hfill\square$
	
	\setcounter{equation}{0}
	\section{Proof of Proposition 2}\label{appendicesB}
	The objective function in the problem $\mathbf{L}\mbox{-}\mathbf{w2}$ can be determined is strictly convex with respect to $\textbf{W}$, since $\textbf{Tr}(\textbf{w} \textbf{w}^H)=|| \textbf{w}||_2^2 $ is a convex function. Constraints (41) and (43a) are all affine. Therefore, Proposition 2 follows. $\hfill\square$ 
	
	\setcounter{equation}{0}
	\section{Proof of Proposition 3}\label{appendicesC}
	We first take the first-order derivatives of $\frac{1}{\theta(\textbf{S})}$, $\gamma(\textbf{S})$, $2^{\eta_1(\textbf{S})}$, and $2^{\eta_2(\textbf{S})}$ as follows: \renewcommand\theequation{C.\arabic{equation}}
	\begin{equation}
		\begin{aligned} \label{3-1}
			(\frac{1}{\theta(\textbf{S})})^{'} =
			&-\frac{1}{(\theta(\textbf{S}))^2} (-\mathbf{A}_s^H \Omega^{-1} \mathbf{g}_n^{t2}(\mathbf{g}_n^{t2})^H \Omega^{-1} \mathbf{A}_s\\
			& -\mathbf{A}_c^H \Omega^{-1} \mathbf{g}_n^{t2}(\mathbf{g}_n^{t2})^H \Omega^{-1} \mathbf{A}_c  ) ,
		\end{aligned}
	\end{equation}
	\begin{equation}
		\small
		\begin{aligned} \label{3-2}
			(\gamma(\textbf{S}))^{'} = \frac{2 \tau B^{\text{UAV}}}{(\textbf{Y}^H \textbf{S} \textbf{Y} + Z)^2}\big((\textbf{Y}^H \textbf{S} \textbf{Y} + Z)\textbf{X}\textbf{X}^H -(\textbf{X}^H \textbf{S} \textbf{X})\textbf{Y}\textbf{Y}^H \big),
		\end{aligned}
	\end{equation}
	\begin{equation}
		\begin{aligned} \label{3-3}
			(2^{\eta_1(\textbf{S})})^{'} = 2^{\eta_1(\textbf{S})} \ln2 \frac{-\sum_{n=1}^{N} D_{n}\frac{T_{11} C^{\text{UAV}} \delta}{f^\text{UAV} 2 \tau}\frac{(\gamma(\textbf{S}))^{'}}{1+\gamma}}{(T_{11}+ \frac{T_{11} C^{\text{UAV}} \delta}{f^{\text{UAV}} 2 \tau} \log(1+\gamma(\textbf{S})))^2}, 
		\end{aligned}
	\end{equation}
	\begin{equation}
		\begin{aligned} \label{3-4}      
			(2^{\eta_2(\textbf{S})})^{'} = 2^{\eta_2(\textbf{S})} \ln2 \frac{-\sum_{\Theta(m) > \Theta(n)} D_{m}\frac{T_{11} C^{\text{UAV}} \delta}{f^\text{UAV} 2 \tau}\frac{(\gamma(\textbf{S}))^{'}}{1+\gamma}}{(T_{11}+ \frac{T_{11} C^{\text{UAV}} \delta}{f^{\text{UAV}} 2 \tau} \log(1+\gamma(\textbf{S})))^2}.
		\end{aligned}
	\end{equation}
	Then, we define constraints (24) and (34a) as $f_4(\textbf{S}_0)=\sum_{n=1}^{N}D_n +\alpha T_{11} \psi(\textbf{S}_0)$ and $f_1(\textbf{S}_0)=\frac{1}{\theta(\textbf{S}_0)} (2^{\eta_1(\textbf{S}_0)} - 2^{\eta_2(\textbf{S}_0)})$, by performing the first-order Taylor expansion on $f_4(\textbf{S}_0)$ and $f_1(\textbf{S}_0)$ 
	, we obtain the following approximation
	\begin{equation}
		\begin{aligned} \label{3-7}
			f_4(\textbf{S}_0) \approx f_4(\textbf{S}^{(k)}) + \bigtriangledown f_4(\textbf{S}^{(k)}) (\textbf{S}_0-\textbf{S}^{(k)}),
		\end{aligned}
	\end{equation}
	\begin{equation}
		\begin{aligned} \label{3-5}
			f_1(\textbf{S}_0) \approx f_1(\textbf{S}^{(k)}) + \bigtriangledown f_1(\textbf{S}^{(k)}) (\textbf{S}_0-\textbf{S}^{(k)}),
		\end{aligned}
	\end{equation}
	where $\bigtriangledown f_4(\textbf{S}^{(k)}) = \alpha T_{11} \frac{1}{1+\gamma(\textbf{S}^{(k)})} (\gamma(\textbf{S}^{(k)}))^{'}$, and $\bigtriangledown f_1(\textbf{S}^{(k)}) = (\frac{1}{\theta(\textbf{S}^{(k)})})^{'}(2^{\eta_1(\textbf{S}^{(k)})} - 2^{\eta_2(\textbf{S}^{(k)})})+\frac{1}{\theta(\textbf{S}^{(k)})}(2^{\eta_1(\textbf{S}^{(k)})} - 2^{\eta_2(\textbf{S}^{(k)})})^{'}$. 
	
	Next, we denote $f_2(\textbf{S}_0)=\frac{1}{\theta(\textbf{S}_0)} (2^{\eta_1(\textbf{S}_0)} - 2^{\eta_2(\textbf{S}_0)}) (T_{11}+\frac{T_{11} C^{\text{UAV}} \delta}{f^\text{UAV} 2 \tau} \log(1+\gamma(\textbf{S}_0)))$. We take the first-order derivative of $f_2(\textbf{S}_0)$ as follows:
	\begin{equation}
		\small
		\begin{aligned} \label{3-6}
			(f_2(\textbf{S}_0))^{'}& = (\frac{1}{\theta(\textbf{S}_0)})^{'} (2^{\eta_1(\textbf{S}_0)} - 2^{\eta_2(\textbf{S}_0)}) (T_{11}+\frac{T_{11} C^{\text{UAV}} \delta}{f^\text{UAV} 2 \tau} f_3(\textbf{S}_0)) \\
			&+ \frac{1}{\theta(\textbf{S}_0)} (2^{\eta_1(\textbf{S}_0)} - 2^{\eta_2(\textbf{S}_0)})^{'} (T_{11}+\frac{T_{11} C^{\text{UAV}} \delta}{f^\text{UAV} 2 \tau} f_3(\textbf{S}_0))\\
			& + \frac{1}{\theta(\textbf{S}_0)} (2^{\eta_1(\textbf{S}_0)} - 2^{\eta_2(\textbf{S}_0)}) (T_{11}+\frac{T_{11} C^{\text{UAV}} \delta}{f^\text{UAV} 2 \tau} f_3(\textbf{S}_0))^{'},
		\end{aligned}
	\end{equation}
	where $(f_3(\textbf{S}_0))^{'}=(\log(1+\gamma(\textbf{S}_0)))^{'} = \frac{(\gamma(\textbf{S}_0))^{'}}{1+\gamma(\textbf{S}_0)}$. By performing the first-order Taylor expansion on $E^{\text{tot}}$, we obtain the following approximation
	\begin{equation}
		\small
		\begin{aligned}
			\label{rew-objfun}
			E^{\text{tot}} = & \sum_{n=1}^{N} f_2(\textbf{S}_0) +\text{Tr}(\textbf{S}_0) T_{11} +\omega^\text{UAV} T_{11} C^{\text{UAV}} (f^\text{UAV})^2 \frac{\delta}{2 \tau} f_3(\textbf{S}_0) \\
			&+\text{Tr}(\textbf{w} \textbf{w}^H) T_2 +\frac{1}{2}v w^{\text{UAV}} (T_{11} +T_2 +\frac{T_{11} C^{\text{UAV}} \delta}{f^\text{UAV} 2 \tau} f_3(\textbf{S}_0) )  \\
			& \approx E^{\text{tot}}(\textbf{S}^{(k)}) + \bigtriangledown E^{\text{tot}}(\textbf{S}^{(k)}) (\textbf{S}_0-\textbf{S}^{(k)}),
		\end{aligned}
	\end{equation}
	where $\bigtriangledown E^{\text{tot}}(\textbf{S}^{(k)}) = \sum_{n=1}^{N} (f_2(\textbf{S}_0))^{'} + T_{11} + \omega^\text{UAV} T_{11} C^{\text{UAV}} (f^\text{UAV})^2 \frac{\delta}{2 \tau} (f_3(\textbf{S}_0))^{'} + \frac{1}{2}v w^{\text{UAV}} \frac{T_{11} C^{\text{UAV}} \delta}{f^\text{UAV} 2 \tau} (f_3(\textbf{S}_0))^{'} $. Therefore, Proposition 3 is proved. $\hfill\square$
	
\end{appendices}

\bibliographystyle{IEEEtran} 
\bibliography{reference}

\vspace{11pt}


\vfill

\end{document}